\def\figcond{1}     
\def\figa#1{%
\ifnum\figcond>0
 \centerline{\hbox{\hsize=4.875in\hss\vbox to 3.45in{\vss%
 \vskip -0.15in%
 \centerline{\hskip  0.10in\epsfig{file=#1.ps,width=5.7in}}
 \vss}\hss}}
\else
 \centerline{\hbox{\hsize=4.875in\hss\vbox to 3.45in{\vss%
 \centerline{Figures or Hardcopys available from koepf@josiah.tau.ac.il}%
 \vss}\hss}}
\fi}
\def\figb#1{%
\ifnum\figcond>0
 \centerline{\hbox{\hsize=5.500in\hss\vbox to 1.90in{\vss%
 \vskip  5.30in%
 \centerline{\hskip  2.00in\epsfig{file=#1.ps,width=7.5in}}
 \vss}\hss}}
\else
 \centerline{\hbox{\hsize=5.500in\hss\vbox to 1.90in{\vss%
 \centerline{Figures or Hardcopys available from koepf@josiah.tau.ac.il}%
 \vss}\hss}}
\fi}
\def\figc#1{%
\ifnum\figcond>0
 \centerline{\hbox{\hsize=6.000in\hss\vbox to 4.30in{\vss%
 \vskip  0.75in%
 \centerline{\hskip -0.75in\epsfig{file=#1.ps,width=7.5in}}
 \vss}\hss}}
\else
 \centerline{\hbox{\hsize=6.000in\hss\vbox to 7.70in{\vss%
 \centerline{Figures or Hardcopys available from koepf@josiah.tau.ac.il}%
 \vss}\hss}}
\fi}
\def\figd#1{%
\ifnum\figcond>0
 \centerline{\hbox{\hsize=4.875in\hss\vbox to 6.30in{\vss%
 \vskip  0.30in%
 \centerline{\hskip  0.10in\epsfig{file=#1.ps,width=5.7in}}
 \vss}\hss}}
\else
 \centerline{\hbox{\hsize=4.875in\hss\vbox to 6.10in{\vss%
 \centerline{Figures or Hardcopys available from koepf@josiah.tau.ac.il}%
 \vss}\hss}}
\fi}
\def\fige#1{%
\ifnum\figcond>0
 \centerline{\hbox{\hsize=4.875in\hss\vbox to 7.50in{\vss%
 \vskip -0.15in%
 \centerline{\hskip  0.10in\epsfig{file=#1.ps,width=5.7in}}
 \vss}\hss}}
\else
 \centerline{\hbox{\hsize=4.875in\hss\vbox to 3.45in{\vss%
 \centerline{Figures or Hardcopys available from koepf@josiah.tau.ac.il}%
 \vss}\hss}}
\fi}
\def\figf#1{%
\ifnum\figcond>0
 \centerline{\hbox{\hsize=4.875in\hss\vbox to 6.60in{\vss%
 \vskip  0.20in%
 \centerline{\hskip  0.10in\epsfig{file=#1.ps,width=5.7in}}
 \vss}\hss}}
\else
 \centerline{\hbox{\hsize=4.875in\hss\vbox to 6.10in{\vss%
 \centerline{Figures or Hardcopys available from koepf@josiah.tau.ac.il}%
 \vss}\hss}}
\fi}
\ifnum\figcond>0
\documentstyle[preprint,floats,aps,epsfig]{revtex}
\else
\documentstyle[preprint,floats,aps]{revtex}
\fi
\begin{document}

\preprint{TAUP-2290-95, hep-ph/9509311}
\draft

\title{Hard diffractive electroproduction of vector mesons in QCD}

\author{Leonid Frankfurt\thanks{\noindent On leave of
    absence from the St.Petersburg Nuclear Physics Institute, Russia.}
    and Werner Koepf\thanks{\noindent Now at Ohio State University, Columbus,
    Ohio}\\ 
  School of Physics and Astronomy\\ 
  Raymond and Beverly Sackler Faculty of Exact Sciences\\
  Tel Aviv University, Tel Aviv, Israel
\\[0.3cm]
Mark Strikman\thanks{\noindent Also  St.Petersburg Nuclear
 Physics Institute, Russia.}\\
Pennsylvania State University, University Park, Pennsylvania
}

\date{Revised May 28, 1996}

\maketitle
\vskip 0.2in
\centerline{{\it in print by\/} Phys.~Rev.~{\bf D}}
\vskip -0.2in

\begin{abstract}
Hard diffractive electroproduction of longitudinally polarized vector mesons 
is calculated within the leading $\alpha_s\ln{Q^2\over\Lambda_{QCD}^2}$ 
approximation of QCD using the leading order parton densities within the 
nucleon.  Novel QCD features of the production of excited states 
and of the restoration of flavor symmetry are analyzed.  At the onset of the 
asymptotic regime, our analysis finds an important role of quark Fermi motion 
within the diffractively produced vector mesons,
and we suggest to use this effect to measure the high momentum tail of the
wave function of the vector mesons.  
We deduce a kinematical boundary for the region of applicability of 
the decomposition of the hard amplitudes over powers of $Q^2$ 
and/or a limit on the increase of the cross sections of hard processes at 
small $x$, and briefly analyze its consequences. We also estimate
the nuclear attenuation of the diffractive electroproduction of vector 
mesons and compare with estimates of the
shadowing of the longitudinal structure function.
\end{abstract}

\newpage

\section{Introduction}

        Diffractive (coherent) production of hadron states in deep inelastic 
lepton-nucleon scattering is a new kind of hard process calculable 
in QCD. It provides the unique possibility to study the properties of vacuum 
exchange in QCD as well as the interplay between soft (nonperturbative) and 
hard (perturbative) contributions to the Pomeron. 
One of the main aimes of such research is to obtain a three dimensional image 
of a hadron as compared to the one dimensional images extracted from the usual 
deep inelastic scattering off hadrons and nuclei. 
The idea to use diffractive electroproduction 
of vector mesons to investigate color coherence effects has been suggested 
many years ago in Refs. \cite{FS88,BM}.

        It has been shown in Ref. \cite{FS89} that the cross sections of hard
diffractive processes off a proton target are proportional to the square
of the gluon distribution in the proton. 
Recently, Brodsky et al. \cite{Brod94} derived within the leading 
$\alpha_s \ln {Q^2\over  \Lambda_{QCD}^2} \ln {1 \over x}$ approximation a
QCD prediction for the diffractive production of vector mesons built 
of light flavors in deep inelastic scattering. Diffractive photo- and 
electroproduction of $J/\Psi$ mesons has been calculated by M.Ryskin 
\cite{Ryskin} within the BFKL approximation of 
pQCD and within the charmonium constituent quark model approximation for the 
light-cone wave function of the $J/\Psi$ meson. 

        On the experimental side, first HERA 
results on $\rho$-meson production at $Q^2 \ge 8$ GeV$^2$ \cite{ZEUS} and on 
photoproduction of $J/\Psi$ mesons \cite{H1,ZEUS1} have confirmed the fast 
increase of the cross section with energy predicted by pQCD.

        In this paper, we extend the QCD analysis of Ref. \cite{Brod94} 
in several directions. 
We explain that the formulae of Ref. \cite{Brod94} are valid within the
conventional leading
$\alpha_s \ln {Q^2\over  \Lambda_{QCD}^2}$ approximation.
So the cross section for hard diffractive electroproduction of vector
mesons is expressed
through the  conventional leading $\ln {Q^2\over\Lambda_{QCD}^2}$
order gluon distribution,
and not by means of its
asymptotics within leading order in $\alpha_s \ln {Q^2\over \Lambda_{QCD}^2} 
\ln {1 \over x}$ as in Ref. \cite{Brod94}. 
This calculation justifies using the LO QCD improved 
parton model densities to compare QCD predictions for the 
electroproduction of vector mesons with experimental data.

        We analyze the average transverse sizes of the $q \bar q$ components 
effective in $\sigma_{\gamma^*_Lp}$, $b_{\sigma_L}$, and in 
$\sigma_{\gamma^*_Lp \rightarrow Vp}$, $b_{\gamma^*_L\rightarrow V}$, and find 
them to decrease strongly with increasing $Q^2$. At $Q^2 \sim 10$ GeV$^2$, 
those sizes are comparatively close, 
$b_{\sigma_L} \sim b_{\gamma^*_L\rightarrow V} \sim 0.3$ fm, 
while at larger $Q^2$ the rate of decrease of this average size is larger 
for $b_{\sigma_L}$. Thus we argue that the current 
$Q^2 \sim 10$ GeV$^2$ data suggest 
that the measured radius of the color distribution within a meson 
is $\sqrt{{3\over 2}}\,b_{\gamma^*_L\rightarrow V} \gtrsim 
0.3 \div 0.4$ fm.   
This observation justifies the application of pQCD for the calculation of
hard diffractive electroproduction of longitudinally polarized
vector mesons. Really, the screening
of the color charges insures the smallness of the effective 
interaction due to asymptotic freedom.
In the case of $\sigma_{\gamma^*_Tp}$, the effective transverse size of the
$q \bar q$ components is expected to be intermediate between those in
$\sigma_{\gamma^*_Lp}$ and those in photoproduction
because the contribution of soft QCD is enhanced for $\sigma_{\gamma^*_Tp}$.
Also, the significant difference between $b_{\sigma_L}$ and
$b_{\gamma^*_L\rightarrow V}$ indicates that substantial next-to-leading order
corrections should be present in diffractive vector meson photo-
and electroproduction which, in turn, should lead to a change of the scale
effective in the gluon density.

        We show that the QCD prediction for the $Q^2$ dependence of the 
$\rho$-meson production cross section is substantially weaker than that 
given by the naive dimensional estimate of 
$\sigma_{\gamma^*p \rightarrow Vp}(x,Q^2) \propto Q^{-6}$ 
as a result of the increase of the parton
distributions due to $Q^2$ evolution \cite{Halina}. We find also a
further slowing down of the $Q^2$ dependence  due to the
transverse quark motion in the vector mesons.
Account of both effects resolves, to a large  
extent, seeming contradictions of predictions of Ref. \cite{Brod94} with 
the $Q^2$ dependence of the cross section of the ZEUS data \cite{ZEUS}.
Thus, investigation of the $Q^2$ dependence of the cross section of 
electroproduction of vector mesons may turn out to be an effective method 
to probe the three dimensional distribution of color in hadrons.

        We analyze the energy and $Q^2$ dependence of the restoration of 
$SU(3)$ flavor symmetry in the electroproduction of vector mesons built of 
different flavors. We find that the ratio of the yields of $\phi$- and 
$\rho$-meson production should increase with increasing $Q^2$ and/or $1/x$. 
We explain also that, in hard diffractive processes, $SU(4)$ flavor
symmetry should be violated in the opposite way
as compared to the low energy regime -- production of heavy flavors should be
enhanced at sufficiently large $Q^2$.
To analyze the onset of the restoration of $SU(4)$ flavor symmetry, we
evaluate diffractive photo- and electroproduction of 
$J/\Psi$ and $\Psi'$ mesons 
within the charmonium model approximation to the light-cone
wave function of the $J/\Psi$ and $\Psi'$
mesons, as suggested in Ref. \cite{Ryskin}, 
and find that account of the quark Fermi motion leads to a significant 
additional suppression of $J/\Psi$ photoproduction by a factor $\sim 1/8$
as compared to the asymptotic estimate, and to a slowing 
down of the onset of the asymptotic regime.
Comparison with the data of Ref. \cite{e401,e516,ZEUS1,prelim} on
$J/\Psi$ photoproduction shows that the pQCD 
predictions can be brought to agree with the
data within the uncertainties in the gluon distribution of a nucleon.
We explain also that the charmonium models differ from QCD 
for large $Q^2$.
We furthermore present estimates for the yields of excited vector meson 
states in diffractive processes, $\gamma^*_L +p \rightarrow V' +p$. The 
interesting new effect, specific for QCD, is that the relative yield of 
excited 
states should increase with increasing $\frac{1}{x}$ and/or $Q^2$.

        Towards the end of the paper, we estimate the boundary of the
kinematical region where, at very small $x$, the fast increase of the 
cross section with decreasing $x$ should slow down 
and where existing methods of hard pQCD seem to be insufficient. We deduce the 
constraint on the kinematical range of applicability of the leading logarithm
approximations and/or on the increase of the cross sections of hard processes
with energy.  We discuss practical consequences of this
constraint and present model estimates for nuclear shadowing of the gluon 
distribution and of the diffractive vector meson electroproduction in 
interactions with heavy nuclei. These effects should be significant in  
kinematics which can be probed at colliding electron-nuclear beams
at HERA and in the LHC heavy ion experiments.

        Diffractive photo- and electroproduction of vector mesons has been
investigated recently in Refs. \cite{kopel2,nemch,nemch2,benha,benha2}
within the constituent quark model based on the 
two-gluon-exchange approximation \cite{low}.  Some of the predictions of 
that model -- like the decrease of the effective size of the vector mesons 
with increasing $Q^2$ \cite{kopel2} or a significant production 
of $\rho'$ mesons \cite{nemch} -- resemble results obtained in this paper.
Other predictions -- as the dip in the $\rho'$ meson production
at small $Q^2$ suggested in Ref. \cite{nemch} 
or the qualitative difference between the behavior of radial and orbital 
(e.g. $D$-state) excitations 
-- do not follow from the QCD formulae of Ref. \cite{Brod94}
and the current paper.

At the same time, we want to emphasize that at sufficiently large $Q^2$, 
where a QCD calculation can be
substantiated, the cross section should be expressed through the distribution
of bare quarks in the vector meson and the parton distributions within the
target \cite{Brod94} but not through the distribution of constituent quarks
as in the models of Refs. \cite{Ryskin,kopel2,nemch,nemch2,benha,benha2}.
There exists no direct relationship between the minimal Fock component of 
the light-cone hadron wave function, which enters in the QCD formulae 
for hard processes, and the wave functions of the constituent quark model.
For example, the minimal
Fock component of the quark-gluon wave function of vector mesons should
decrease at large $k_t^2$ as a power of $k_t^2$ (cf. Ref. \cite{BL})
but not like a Gaussian as used in Refs. 
\cite{kopel2,nemch,nemch2,benha,benha2}.  
Just this difference is responsible for relevant effects found in 
the current work
which are due to the transverse quark Fermi motion in the diffractively 
produced vector mesons. The qualitative difference between QCD calculations 
and the two-gluon-exchange model of Ref. \cite{low} has been discussed
in Ref. \cite{Brod94}, and additional references to other works in that realm 
can also be found there.

        The organization of this work is as follows: In Sect. II, we give
leading $\alpha_s \ln {Q^2\over \Lambda_{QCD}^2}$ predictions for the
diffractive vector meson production in deep inelastic scattering, and, in
Sect. III, we justify our pQCD approach listing the necessary approximations.
In Sect. IV, we give estimates of the effects of transverse quark motion
within the diffractively produced vector mesons
and we discuss the NLO $Q^2$ rescaling. In 
Sect. V, we compare
our results with the few currently available data. In Sect. VI, we discuss
the restoration of flavor symmetry and, in Sect. VII, the production of excited
states. In Sect. VIII, we estimate a kinematical constraint on the 
region of applicability of the leading logarithm approximations, and, in Sect. 
IX, nuclear shadowing 
in the hard diffractive production of vector mesons as
well as the longitudinal structure function of a nucleus are considered. 

\section{Leading $\alpha_s \ln{Q^2 \over \Lambda_{QCD}^2}$
predictions for the  diffractive vector meson production 
in DIS}

It was demonstrated in Ref. \cite{Brod94},
that a new kind of hard process, the coherent electroproduction of 
vector mesons off a target T,
\begin{equation}
\gamma^* + T \rightarrow V+T \ ,
\label{eq1}
\end{equation}
is  calculable in QCD for $Q^2\gg M_V^2$ within the leading   
$\alpha_s \ln{Q^2 \over \Lambda_{QCD}^2}\ln {1 \over  x}$ 
approximation.
Here, $V$ denotes any vector meson ($\rho,\omega,\phi, J/\Psi$) or
its excited states.  For comparison with previous works and appropriate 
references see Ref. \cite{Brod94}.

      The idea behind the calculation of hard diffractive processes is that,
when the coherence length, $l_{c}= {1 \over 2 m_Nx}$, exceeds the diameter of
the target, $2 r_T$, the virtual photon transforms into a hadron component well
before reaching the target and the final vector meson $V$ is formed
well past the target. The hadronic configuration of the final state is
then the result of a coherent superposition of all those hadronic
fluctuations, $|n\rangle$, in the photon wave function whose masses 
satisfy the condition:
\begin{equation}
-{t_{min} r_{T}^2\over 3} \ll 1 \ .
\label{eq2}
\end{equation}
Here, $t_{min}$ is the minimal momentum transferred to the target:
\begin{equation}
-t_{min}= \left({M_n^2+Q^2 \over 2 q_0 } \right)^{\!2} = 
\left(1 +{M_n^2  \over Q^2 } \right)^{\!2} m_N^2x^2 \ .
\end{equation}

  Thus, as in the more familiar leading twist deep inelastic processes, the
calculation should take into account all possible diffractively produced
intermediate hadronic states satisfying Eq. (\ref{eq2}).  The use of
completeness over those states
allows us to express the result in terms of quark and 
gluon distributions, as in the
case of other hard processes.  The matrix element of electroproduction
of a vector meson can thus be written as a convolution of the light-cone
wave function of the photon, $\psi_{\gamma^* \rightarrow |n\rangle}$,
the scattering amplitude of the hadron state, $~A(nT)$,
and the wave function of the vector meson, $\psi_{V}$:
\begin{equation}
A= \psi^\dagger_{\gamma^*  \rightarrow |n\rangle}  \otimes
A (nT) \otimes \psi_{V} \ .
\label{eq3}
\end{equation}
In the case of a longitudinally polarized photon with sufficiently
large $Q^2$, the sum over
intermediate states $|n\rangle$  is  a $q\bar q$ pair.
This can be demonstrated by transition into impact parameter space 
and by showing that 
the essential distances, $b$, between the quarks in the wave function of 
the photon are 
$b^2\propto 1/Q^2$.  
The situation is qualitatively different, however, for a
transversely polarized photon due to the singular behavior of the vertex
$\gamma^*_T \rightarrow q\bar q$ when one of the partons carries a small
fraction of the photon momentum. In this case, soft
and hard physics compete in a wide range of $Q^2$ and $x$.
  
Within the leading logarithmic approximation, i.e.,
$\alpha_s  \ln {Q^2\over  \Lambda_{QCD}^2} \ln {1 \over  x} 
\sim 1$, when terms of the order
$\alpha_s \ln {Q^2\over  \Lambda_{QCD}^2}$ 
are neglected, the final result \cite{Brod94} for the cross section 
for the production of longitudinally polarized vector meson states, when 
the momentum transferred to the target $t$ tends to zero, 
is\footnote{In Ref. \cite{Brod94}, a factor 4 was missed in the numerator 
of Eq. (\ref{eq5}) \cite{errata}. We are indepted to Z. Chan for pointing 
this out (A.Mueller private communication).}
\begin{equation}
\left. {d\sigma^L_{\gamma^*N\rightarrow VN}\over dt}\right|_{t=0} =
{12\pi^3\Gamma_{V \rightarrow e^{+}e^-} M_{V}\alpha_s^2(Q)\eta^2_V
\left|\left(1 + i{\pi\over2}{d \over d\ln x}\right)xG_T(x,Q^2)\right|^2
\over \alpha_{EM}Q^6N_c^2} \ .
\label{eq5}
\end{equation}
Here,
$\Gamma_{V \rightarrow e^{+}e^-}$ is the decay width of the vector
meson into an $e^+e^-$ pair. The parameter $\eta_V$ is defined as
\begin{equation}
\eta_V\equiv {1\over 2}{\int{dz\,d^2k_t\over z(1-z)}\,\Phi_V(z,k_t)\over
\int dz\,d^2k_t\,\Phi_V(z,k_t)}  \ ,
\label{eq6}
\end{equation}
where $\Phi_V$ is the light-cone wave function of the 
$q \bar q$ component of the vector meson.

The factor $|\alpha_s(Q^2) x G_T(x,Q^2)|^2$ in Eq. (\ref{eq5}) originates 
from the
square of the cross section for the  high-energy interaction of a small-size
$q\bar q$ configuration with any target. It  can be unambiguously calculated 
in QCD for low $x$ processes by applying the QCD factorization theorem.  In 
the approximation where only the leading $\alpha_s  \ln {Q^2\over  
\Lambda_{QCD}^2}$ terms are accounted for, the result is 
\cite{BBFS93,FMS93}
\begin{equation}
\sigma(b^2)={\pi^2\over 3}\left[ b^2\alpha_s(Q^2)xG_{T}
(x,Q^2)\right]_{x=Q^2/s, Q^2=\lambda/b^2} \ ,
\label{eq7}
\end{equation}
where $b$ is the transverse distance between the quark $q$ and the
antiquark $\bar q$, $xG_T(x,Q^2)$ is the gluon distribution in the
target, and $\lambda \approx 10$ according to our estimate in the
next section. In this equation, the $Q^2$ evolution and small $x$
physics are properly taken into account through the gluon
distribution.
$\sigma(b^2)$ can be inferred from the calculations of Ref. \cite{mueller}, 
which were performed in the leading $\alpha_s \ln x$ approximation of pQCD
using certain input from hadronic quark models.  
Also, a quantity which can be related to $\sigma(b^2)$ 
has been calculated in Ref. \cite{levrys} within the leading
$\alpha_s \ln x$ approximation of pQCD while making 
specific assumptions on the parton structure of the constituent quarks. 

The upper bound of the $x$-range of applicability of Eq. (\ref{eq5}) is
dictated by the condition that the formation time parameter should be
much greater
than the interaction time, which corresponds to the requirement that
the longitudinal coherence length is larger than the target's diameter:
\begin{equation}
x \ll {1 \over 4m_N r_T} \approx 0.06 \ .
\label{cohlen}
\end{equation}
 
       Eqs. (\ref{eq5}) and (\ref{eq7}) can be easily generalized to the 
somewhat less restrictive leading  $\alpha_s \ln{Q^2\over {\Lambda_{QCD}}^2}$ 
approximation.
Derivation of the respective 
formulae is relatively simple for hard diffractive 
processes initiated by longitudinally polarized photons because
longitudinally polarized photons produce quarks in a PLC (point-like 
configuration) only.
It can be demonstrated by transition into impact parameter space that 
the essential distances, $b$, between the
quarks in the wave function of a longitudinally polarized
photon with virtuality $Q^2$ are $b^2 \propto 1/Q^2$.
Using $b$ as the small parameter, it is easy to determine which Feynman 
diagrams dominate for the hard diffractive process. It is important to
note that the integral over the distance between the quarks produced by the 
longitudinally polarized
photon does not lead to terms proportional to $\ln b^2$ in the evaluation of 
the box diagrams. So the whole contribution is concentrated at small $b$.
As a result, the factorization/decoupling theorem of QCD
can be applied to factorize 
the hard part of the amplitude. So, to calculate the amplitude in the leading  
$\alpha_s \ln {Q^2\over  \Lambda_{QCD}^2}$ approximation, we need to 
accurately evaluate the hard blobe and then convolute it with the parton 
distributions. The hard blobe is described by the Feynman diagrams in Figs.
1a and 1b, where gluons are attached to the quark box diagram. 


\begin{figure}[htb]
\figb{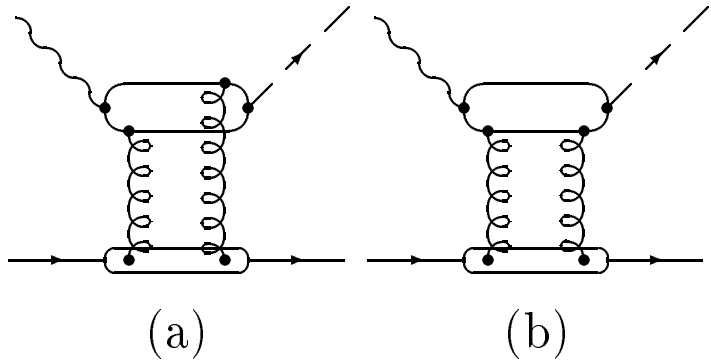}
\vspace{0.2cm}
\caption{Feynman diagrams relevant for the evaluation of the cross 
section of diffractive electroproduction of longitudinally polarized vector 
mesons, i.e., the $\gamma_L^* + T \rightarrow V + T$ process, in leading 
$\alpha_s\ln{Q^2\over {\Lambda_{QCD}^2}}$ approximation.}
\end{figure}
     
    Direct calculation of the sum of Feynman diagrams for the $\gamma_L^* 
\rightarrow V$ transition of Figs. 1a and 1b
where only two gluons couple to the hard blobe 
-- although all interactions of 
these two gluons with the target are included -- leads to seemingly the 
same formulae as in Ref. \cite{Brod94}.
Really it is easy to demonstrate that within the leading
$\alpha_s \ln {b^2 \Lambda^2_{QCD}}$ approximation, which is
equivalent to the leading $\alpha_s \ln {Q^2\over \Lambda^2_{QCD}}$
approximation, the dominant contribution is given by the diagrams with
exchange of longitudinally polarized gluons. The diagrams corresponding to the
exchange of transversely polarized gluons and those due to the quark sea in
the nucleon are suppressed at small $x$.
For a detailed discussion see Ref. \cite{CFS}.

The major difference from \cite{Brod94} is that, within the 
approximation discussed above, the full gluon distribution
in the target calculated within the leading
$\alpha_s \ln{Q^2\over\Lambda_{QCD}^2}$
approximation should be used.
In the formulae of Ref. \cite{Brod94}, however,
$xG_T(x,Q^2)$ is the gluon structure function calculated within the leading  
$\alpha_s \ln {Q^2\over  \Lambda_{QCD}^2}\ln {1 \over x} $
approximation. 
Remember, however, that corrections of powers $(lb)^n$ are numerically 
small and negative \cite{Brod94}.  Here, $l$ is the gluon's transverse 
momentum within the "pomeron".

In principle, the nondiagonal gluon distribution
enters into those formulae.
It will be explained in the next section that the usual parton distribution
in the target $T$ is a reasonable approximation to this
nondiagonal density. For a better understanding of
the definition of the nondiagonal parton densities,
these formulae should be compared to a similar one for the total 
photoabsorption cross section off a target $T$:
\begin{equation}
\sigma_{\gamma^*_LT} 
= \int_x^1  d\alpha \, G_T(\alpha,Q^2) \sigma_{\gamma^*_Lg}(\alpha \nu,Q^2)
\ .
\label{sigmat}
\end{equation}
Eq. (\ref{sigmat}) is valid within the leading 
$\alpha_s\ln {Q^2 \over \Lambda^2_{QCD}}$
approximation, and the $Q^2$ evolution is contained in the $Q^2$ 
dependence of the parton distribution in the target $T$.

The analysis of the Feynman diagrams for $Im A_{\gamma^*_LT \rightarrow VT}$
shows that, for diffractive vector meson production, the region of 
integration over $\alpha$ is restricted as:
\begin{equation}
\alpha = x +{M^2_{q \bar q} \over \nu} \ .
\label{al}
\end{equation}
Here, 
\begin{equation}
M^2_{q \bar q} = {m^2_q +k^2_t \over z(1-z)}
\end{equation}
is the invariant mass of the $q \bar q$ pair,
and $z$ and $k_t$ define the light-cone 
momentum carried by the quark/antiquark in the 
wave function of the $\gamma^*_L$. The second term in Eq. (\ref{al})
can be neglected for the production of vector mesons
built of light quarks, since effectively $k_t^2 \ll Q^2$.
This is because the wave function of the vector mesons 
decreases comparatively fastly 
with $k_t^2$ up to the onset of the pQCD regime.  This 
explains why it is a reasonable approximation in 
$Im A_{\gamma^*_LT \rightarrow VT}$ -- but 
not in Eq. (\ref{sigmat}) -- to pull the parton densities 
at $\alpha \approx x$ out of the integrand. 

The respective
hard cross sections can be easily calculated through the miminal 
Fock components of the wave function of the vector meson.
Thus, we deduce our asymptotic expression for the cross section of the 
electroproduction of vector mesons:
\begin{equation}
\left. {d\sigma^L_{\gamma^*N\rightarrow VN}\over dt}\right|_{t=0} =
{12\pi^3\Gamma_{V \rightarrow e^{+}e^-} M_{V}\alpha_s^2(Q)T(Q^2)\eta^2_V
\left|\left(1 + i{\pi\over2}{d \over d\ln x}\right)xG_T(x,Q^2)\right|^2
\over \alpha_{EM}Q^6N_c^2} \ .
\label{eq5a}
\end{equation}
The factor $T(Q^2)$ accounts for preasymptotic effects, i.e.,
$T(Q^2 \rightarrow \infty)=1$,
see the discussion in Sect. IV.

 The major practical  
difference between the formulas deduced within the leading   
$\alpha_s \ln{Q^2\over  \Lambda_{QCD}^2}\ln{1 \over x}$ and the leading 
$\alpha_s \ln{Q^2\over  \Lambda_{QCD}^2}$ approximation (i.e., within the
evolution equation) is that, in the leading $\alpha_s \ln{Q^2 \over 
\Lambda_{QCD}^2}$ approximation, 
$xG_T(x,Q^2)$ should be interpreted as the conventional parton distribution. 

\section{Justification of the pQCD approach for diffractive vector 
meson electroproduction in DIS}

        The formal derivation of the pQCD expressions for the cross sections of 
hard diffractive processes in terms of the parton distributions in the target
consists of several steps. The first is to demonstrate that, at small $x$,
it is legitimate to use completeness over hadronic states to express the
amplitude in terms of hard interactions of quarks and gluons. We discussed 
this in the previous subsection. The second step is to prove that the process 
is really hard and that it is legitimate to apply the QCD factorization 
theorem, and the third step is to show that the nondiagonal parton 
distributions of the target can be approximated through the conventional 
parton densities.

\subsection{Applicability of pQCD and small transverse distances}

        To better understand the issue of applicability of pQCD to the process 
(\ref{eq1}), it is convenient to perform the Fourier transform of the amplitude
into the impact parameter space of the $q \bar q$ pair, which leads to 
\begin{equation}
{\cal A}_{\gamma^*_LN\rightarrow VN}
\propto \int d^2b~dz~\psi^\dagger_{\gamma^*_L}(z,b)~\sigma(b^2)~\psi_{V}(z,b)
\ .
\label{brep}
\end{equation}
Here, the Sudakov variable $z$ denotes the fraction of the photon momentum 
carried by one of the quarks, $b$ is the transverse distance between the quark 
and antiquark within the photon, and 
$\sigma(b^2)={\pi^2\over 3} b^2 \alpha_s(b^2) xG_N(x,b^2)$ is the color-dipole 
cross section of Eq. (\ref{eq7}) off a nucleon target. 
The advantages of the description of high energy processes in terms 
of light-cone wave 
functions of bound states in the impact parameter representation 
has been understood long ago (cf. H.Cheng and T.T.Wu \cite{Wu} and references 
therein).  The 
$q\bar q$ component of the light-cone wave function of 
a longitudinally polarized virtual 
photon reads 
\begin{equation}
\psi_{\gamma_L^*}(z,b) = 2\,Q\,z(1-z)\,K_0\left(Qb\sqrt{z(1-z)}\right)
\label{blong}
\end{equation}
in the limit of vanishing quark mass, and
where $K_0$ is the Hankel function of an imaginary argument.

        To estimate which values of $b$ dominate in the integral, in Eq. 
(\ref{brep}) we approximate $\psi_{V}(z,b)$ by 
\begin{equation}
\psi_V(z,b) \propto z(1-z)\,\mu b\,K_1(\mu b)
\ ,
\label{bvecl}
\end{equation}
This form accounts for the common wisdom of the $z$ dependence of the wave 
function of a longitudinally polarized vector meson, while the $k_t$ 
dependence is chosen\footnote{In QCD, the expected behavior of
$\propto {1 \over k_t^2}$ at large $k_t$ corresponds to even smaller interquark
distances in the wave function of the longitudinally polarized photon.}
to be $\propto {1 \over (k_t^2+\mu^2)^2}$.

The parameter 
$\mu$ is related to the average transverse momentum of a quark/antiquark within
the $q \bar q$ component of the meson's wave function through $\sqrt{\left< 
k^2_t\right>} = \mu/\sqrt{2}$. In our numerical analysis, we vary $\sqrt{\left<
k^2_t\right>}$ between 300 and 600 MeV/c. Our guess is that, in reality,
$\sqrt{\left< k^2_t\right>}$ should be of the order of $400 \div 500$ MeV/c,
which is somewhat larger than the partons' average transverse momenta in 
hadrons. This is because, from the whole set of Fock components of the hadron 
wave function, the overlap with the wave function of the $\gamma^*_L$ 
emphasizes the component which has no soft gluon field. An analysis of the 
energy denominators relevant for the wave functions of vector mesons shows that,
in order to avoid a spurious pole, the $q \bar q$ component's energy 
denominator may dominate only in the kinematics
\begin{equation}
{m_q^2 +k_t^2 \over z(1-z)} \gg M_V^2 \ .
\end{equation}
This condition is restrictive only for the minimal Fock space component.
When the number of constituents is large, the energy denominator is large 
more or less automatically.

        Now we need $\alpha_s(b^2) x G_N(x,b^2)$.
We estimate it based on the impact parameter representation for 
$\sigma_L(x,Q^2)$,
\begin{equation}
\sigma_L(x,Q^2) = 
\int d^2b~dz~|\psi_{\gamma_L^*}(z,b)|^2~\sigma_{q \bar q N}(x,b) \ ,
\label{sigmal}
\end{equation}
the nucleon's total longitudinal photoabsorption cross section, where 
\begin{equation}
\sigma_{q \bar q N}(x,b)
 = {\pi^2  \over 3} \alpha_s(Q^2)~b^2 \left. x G_N(x,Q^2) 
\right|_{Q^2 = {\lambda \over b^2}} \ .
\label{sigmabb}
\end{equation}
In the derivation of this equation, to simplify the estimates, we 
again pulled out the parton distributions in the mean point $x$.

        We perform the analysis in two steps. First we use Eq. (\ref{sigmabb}) 
and neglect the dependence of $\alpha_s(b^2)\,x G_N(x,b^2)$ on $b^2$ to 
estimate an average $b(Q)$ for this process. 
The latter we define as the value of $b$ up to which we have to integrate
to saturate 50\% of the integral in Eq. (\ref{sigmal}). 
We then identify $G_N(x,b) \approx G_N(x,b(Q))$, determine a new average
$b(Q)$, and then iterate this procedure to self-consistency.
The final result of $b(Q)$ for $\sigma_L(x,Q^2)$ obtained in
this manner is plotted as a solid curve in Fig. 2. 


\begin{figure}[htb]
\figa{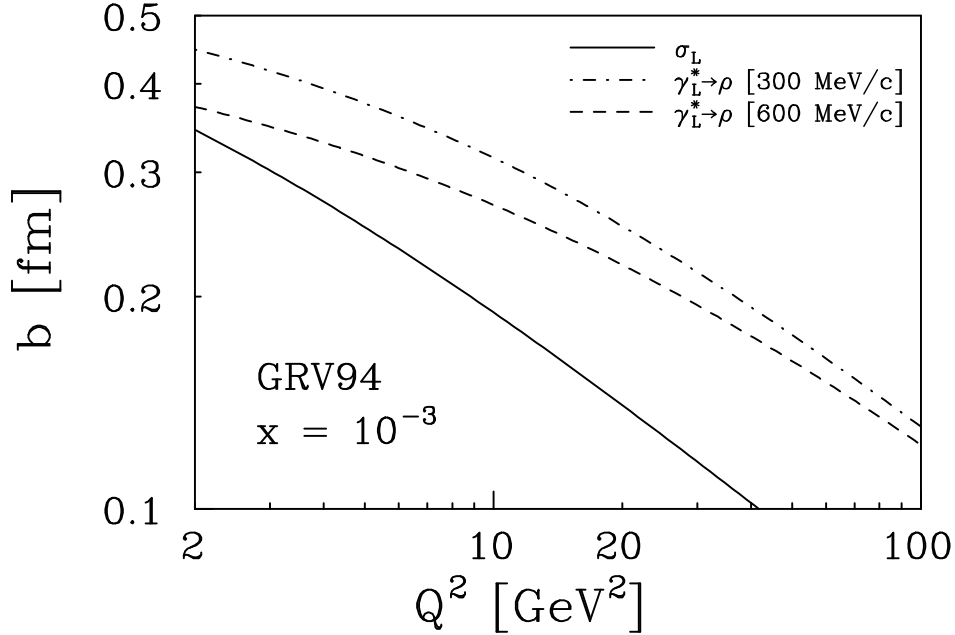}
\caption{Average transverse size of the $q\bar q$ components effective in
${\cal A}_{\gamma^*_LN \rightarrow VN}$ of Eq. (\protect\ref{brep}) and 
$\sigma_L$ of Eq. (\protect\ref{sigmal}). In our numerical calculations,
we used the GRV94 parametrization \protect\cite{GRV} of the nucleon's gluon 
density.}
\end{figure}

Thus, approximately $\lambda \approx 9.2$ or 
\begin{equation}
G_N(x,b^2) \approx G_N\left(x,Q^2={9.2\over b^2}\right) 
\end{equation}
for $x \sim 10^{-3}$. For fixed $Q^2$, the transverse size $b(Q)$ 
slowly decreases with decreasing $x$: between $x=10^{-3}$ and $x=10^{-4}$ 
it drops by about $10\%$.  The above reasoning accounts for some of the next 
to leading order $\alpha_s f\!\left(\alpha_s \ln {Q^2 \over 
\Lambda^2_{QCD}}\right)$ effects, based on the intuitive idea that the 
interaction is determined mostly by the spacial region occupied by color.
Note that a numerically similar result was obtained by Nikolaev and 
Zakharov \cite{nikol} in a consideration of $F_L$ and $\partial F_T/
\partial \ln Q^2$ ($\lambda \sim A_\sigma \approx 10$).  However,
in the derivation of the quantity $A_\sigma$ in Ref. \cite{nikol},
the corresponding integrals were actually dominated by large $b$'s for which 
it is not clear whether the expression for the color-dipole cross section of
Eq. (\ref{eq7}) is at all applicable.
Besides, no $x$-dependence of their $A_\sigma$ was discussed 
in Ref. \cite{nikol}.

        Then we use this value of $b(Q)$ to calculate the average transverse 
distances, $b_{\gamma^*_L \rightarrow \rho}(Q)$, relevant for longitudinal
$\rho$-meson production.  The results of such a calculation for $\sqrt{\left< 
k^2_t\right>} = 300$ and 600 MeV/c are presented also in Fig. 2. One can 
see that $b_{\gamma^*_L \rightarrow \rho}$ is quite small, however its 
numerical value is rather sensitive to the average value of 
$\sqrt{\left< k^2_t\right>}$. We observe also that $b_{\gamma^*_L \rightarrow 
\rho} > b_{\sigma_L}$.  This difference may be rather small for $Q^2 
\lesssim 10$ GeV$^2$, but it becomes significant for $Q^2 \sim 100$ GeV$^2$.
 
        Note that the values we find for $b_{\gamma^*_L \rightarrow \rho}$ are
substantialy smaller than the typical distances between the valence quarks in 
a meson, $2 r_M  \sim 1.2$ fm.  So, color is sufficiently screened within the 
wave function of the produced vector meson, and asymptotic freedom
can be used to justify the applicability of pQCD for the calculation
of diffractive electroproduction of $\rho$-mesons at $Q^2 \ge 10$ GeV$^2$.

        Our results for the elementary color-dipole cross section of a small 
size $q\bar q$ pair with a nucleon target, $\sigma_{q \bar q N}(x,b)$ of Eq. 
(\ref{sigmabb}), are depicted in Fig. 3 for the GRV94 parametrization 
\cite{GRV} of the 
nucleon's gluon density and for various values of $x$. As expected, the cross
section rises approximately quadratically with $b$ and it increases 
dramatically with decreasing $x$. Note, however, that only for the smallest 
$x$, the cross section $\sigma_{q \bar q N}(x,b)$ approaches normal hadronic 
values ($\gtrsim 20$ mb). However, for such $x$, it reaches those already at 
quite small values of $b$, which, in turn, hints a limit of applicability
of the leading logarithm approximations discussed here. 
This will be discussed further in Sect. VIII.


\begin{figure}[htb]
\figa{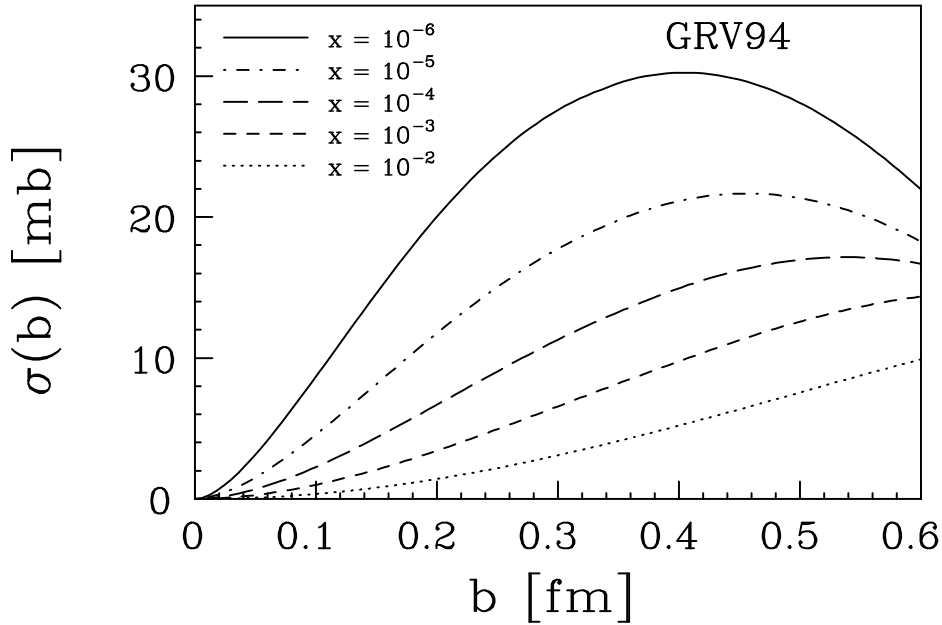}
\caption{Color-dipole cross section, $\sigma_{q \bar q N}(x,b)$ of Eq. 
(\protect\ref{sigmabb}), as a function of the transverse size of the 
$q\bar q$ pair for various values of $x$ and for the GRV94 parametrization
\protect\cite{GRV} of the nucleon's gluon density.}
\end{figure}

Results for an analogous evaluation of the average transverse distances
relevant for the photo- and electroproduction of charmonium mesons are depicted
in Fig. 4. There, we employed nonrelativistic charmonium wave
functions calculated from a power-law \cite{charm1} and a logarithmic
potential \cite{charm2}, which both describe $\Gamma_{J/\Psi \rightarrow
e^+e^-}$ reasonably well, and we set $z={1\over 2}\left(1+{k_z\over 
m_c}\right)$. This yields $\psi_{\gamma^{(*)}\rightarrow c\bar c}(r) \propto
{e^{-\mu r}\over r}$, 
with $\mu=\sqrt{m_c^2+{Q^2\over 4}}$, for the wave function of
the $c\bar c$ component of the (virtual) photon.  Note, however, that
at sufficiently large $Q^2$ -- which is the only domain at which the
formulae for any vector meson production can be consistently proven
in QCD -- $\int d^2k_t\,\psi_{J/\Psi}(k_t,z) \propto z(1-z)$
due to the $Q^2$ evolution \cite{CZ,BL}, which is qualitatively different
from the behavior expected for the nonrelativistic charmonium wave functions.
The question which arises is thus whether hard physics is applicable at 
smaller $Q^2$, and so we are effectively analyzing physics hidden under the 
assumption that the nonrelativistic charmonium wave functions
approximate the light-cone wave function.


\begin{figure}[htb]
\figa{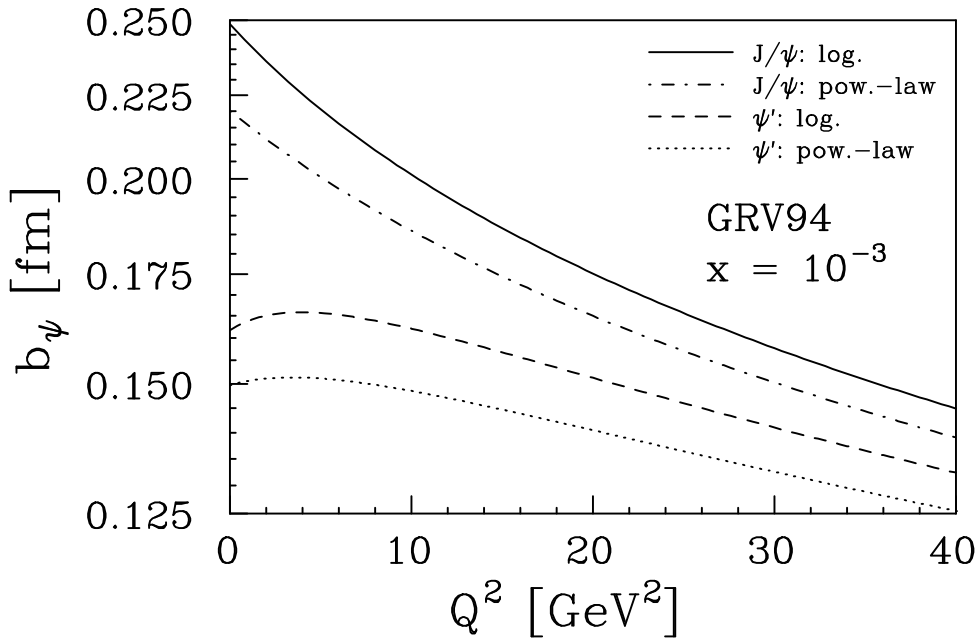}
\caption{Average transverse sizes of the $c\bar c$ components effective in
${\cal A}_{\gamma^{(*)}N \rightarrow V_c N}$ for $J/\Psi$ and $\Psi'$ photo-
and electroproduction.}
\end{figure}

        We observe that for $J/\Psi$ photoproduction the corresponding relevant
distances, $b_\Psi$, are similar
to the average interquark distances in the $J/\Psi$ mesons ($\sim 0.4$ fm), 
while for the photoproduction 
of the excited state, the $\Psi'$, the distances that are most relevant for the
photoproduction process are significantly smaller than the respective average 
interquark separations ($\sim 0.7$ fm). Within the charmonium model 
approximation for the light-cone wave function of the $\Psi'$ meson,
this is due to the nodal structure of
the wave function of the $\Psi'$, as was also observed by Benhar et al. 
\cite{benha} in the framework of the two-gluon-exchange model \cite{low}.
However, for charmonium electroproduction at larger $Q^2$, the distances
relevant in the amplitude are much smaller than the average interquark
separations in the charmonium mesons.

        We can now also estimate the contribution of the hard blobe to the
$t$ dependence of the vector meson electroproduction cross section.
In detail, we fit
\begin{equation}
\int d^2b~dz~\psi_{\gamma^*_L}(z,b)~\sigma(b^2)~\psi_{V}(z,b)~
e^{-iz{\bf p_t}{\bf b}} \propto \left.
e^{B_V t/2}\right|_{t=-{\bf p}_{\bf t}^2}
\label{btdep}
\end{equation}
by an exponential at the vicinity of $t=0$. The slope parameter that
we find, 
$B_V \lesssim 0.7$ 
GeV$^{-2}$, is much smaller than, e.g., the
experimentally observed slope, $B_\rho=4.6 \pm 0.8$ GeV$^{-2}$ \cite{NMC1}, of 
the $\rho$-production cross section off the nucleon. Hence, for all current
practical considerations, the hard blobe does not contribute significantly
to the $t$ dependence of the cross section, which hence originates almost
entirely from the target's two-gluon form factor, $G_{2g}(t)$, and which 
is thus expected to be universal for all hard processes driven by the 
gluon density.

        In the case of a transversely polarized $\gamma^*$, the contribution 
of large $b$, where nonperturbative QCD effects are dominant, is not 
suppressed. Those contributions arise from very asymmetric $q\bar q$
pairs with $z \sim 0$ or $1$ 
\cite{Bj71,FS88,nemch}, 
and, for transverse polarizations, 
in the integrand of the amplitude the regions $z$ or $(1-z)\approx 1/Qb$ thus 
give important contributions, which are usually suppressed 
at large $Q^2$ by a Sudakov type form factor. 
In fact, it is easy to check that, for the case of transversely polarized
photons, the contribution of this regions to the asymptotic expressions
contains a logarithmic divergency. Hence, in this case, the integral over 
transverse momenta does not decouple.

        On the other hand, the increase of $xG_N(x,b)$ at small $x$ works 
towards enhancing the perturbative contribution of small $b$ in this case 
as well. This is because the contribution of quark configurations with small 
$b$ increases faster with decreasing $x$ than the contribution from soft QCD. 
At what $x$ and $Q^2$ the pQCD regime will dominate for reactions initiated 
by a $\gamma^*_T$ is an open question since the contributions from soft QCD 
are not under control in this case. 
Currently, we are nethertheless estimating
the hard, perturbative contribution to $\sigma_{\gamma^*_TN\rightarrow VN}$
\cite{progress}.

Furthermore, since the formulae in this paper were deduced in the leading 
$\alpha_s \ln{Q^2\over \Lambda_{QCD}^2}$ approximation only, there is an 
uncertainty in the theoretical formulae whether the gluon 
distributions should be evaluated at $Q^2$ or say at $Q^2/2$.
A similar problem exists 
for calculations of the leading twist contribution to the total DIS cross 
section within the leading $\alpha_s \ln{Q^2\over \Lambda_{QCD}^2}$ 
approximation.
Note, also, that the substantial difference between $b_{\sigma_L}(Q^2)$
and $b_{\gamma^*_L\rightarrow V}(Q^2)$ which we found in this
subsection indicates that significant next-to-leading order
corrections should be present in diffractive vector meson photo-
and electroproduction.  We will give phenomenological estimates of 
these effects in 
Sects. IV.C and V.

\subsection{The nondiagonal parton distributions}

        In this subsection, we show that the nondiagonal parton distributions 
of the target can be expressed through the diagonal -- conventional -- parton 
densities. The first observation is that the momentum transfered to the 
target, $t_{min} = -m_N^2 x^2 (1+{M_V^2\over Q^2})$, is negligible. 

        A more important issue is that the difference between the masses of 
the virtual photon and the vector meson leads to a difference between the 
light-cone fractions of the target momenta in the initial ($x_i$) and final 
($x_f$) states. To elucidate this point, let us consider diffractive 
vector meson electroproduction in the center of mass of the $\gamma^* +T$ 
system. Suppose the initial target has momentum $P$. Kinematical 
considerations show that the final target has momentum $P(1- (Q^2+M_V^2)/s)$, 
and thus the difference between $x_i$ and $x_f$ is $x$ \cite{Halina}.

        For the hard part of the diffractive amplitude, the dependence on 
fractions of the target momentum carried by the interacting partons in the 
initial and final states is  calculable in pQCD. This simplification arises 
since in the kinematics of diffractive processes, which dominate in the
considered diagrams, it is legitimate to neglect in the propagators of the 
exchanged particles the contribution of the
longitudinal part of their momenta as compared to the contribution of the 
transverse components. This can be easily proven within the 
$\alpha_s \ln{Q^2\over \Lambda_{QCD}^2}$
and/or $\alpha_s \ln{Q^2\over \Lambda_{QCD}^2}\ln {1 \over  x}$
approximation by direct evaluation of the leading Feynman diagrams. 
In reality, this argument is valid beyond the pQCD regime since it explores
specifics of multipheripheral and multiregge kinematics.

        Thus, we conclude that the initial and final parton wave 
functions of the target practically 
coincide, and it is a reasonable numerical approximation
to use the diagonal parton densities. This question will be considered in
detail in Ref. \cite{CFS}.  Here, we have explained that the tiny 
momentum transfered to the target does not influence the hard
blobe, which is specific for small $x$ physics.

\section{Estimates of preasymptotic effects}

\subsection{Vector mesons built of light quarks}

     The calculation discussed above is applicable at suffiently large 
$Q^2$, i.e., only when the transverse momenta in the $q\bar q$ pair
are neglected, can the cross section for electroproduction of longitudinally
polarized vector mesons be written in the form of Eq. (\ref{eq5}).
To analyze the onset of the asymptotic regime, we consider here 
the effects of the Fermi motion of the quarks within the diffractively
produced vector meson.  
As originally suggested in Ref. \cite{hepph}, 
we do this by keeping the quarks' momenta
in the propagator of the virtual photon, as the numerical coefficient
which accompanies the leading term is large ($\approx {32 k_t^2\over Q^2}$).
We will show that account of Fermi motion of the quarks within the vector 
meson leads to a significant suppression of the cross section at moderately 
large $Q^2$. Thus, the formulae for diffractive 
electroproduction of vector 
mesons accounting for this Fermi motion 
contain an additional factor, $T(Q^2)$, which 
can be straightforwardly obtained from Eq. (2.15) of Ref. \cite{Brod94},
and which, within the approximations discussed above, has the form:
\begin{equation}
T(Q^2)=\left({
\int_0^1 dz \int_0^{Q^2}\!d^2k_t \, \psi_V(z,k_t)
\left(-{1\over 4}\Delta_t\right)
\left[{Q^4 \over Q^2 + {k_t^2+m^2\over z(1-z)}}\right]
\over 
\int_0^1 {dz\over z(1-z)}  \int_0^{Q^2}\!d^2k_t \, \psi_V(z,k_t) 
}\right)^{\!\!2} \ .
\label{tfactor}
\end{equation}
Here, $\Delta_t$ is a two dimensional Laplace operator which acts on the
wave function of the photon. We neglect here the current quark mass
of the light quarks, which is small on the scale of the discussed
phenomena.

    To estimate the role of Fermi motion effects, we made estimates 
of the $Q^2$ dependence of $T$ for three types of vector meson wave functions, 
$\psi_V(z,k_t) \sim z(1-z) \phi(k_t)$.  For the nonperturbative piece,
$\phi(k_t)$, we choose either $\phi(k_t) \propto {1\over (k_t^2 + \mu^2)^2}$, 
as in the last section (labeled "dip."), or $\phi(k_t) \propto \exp(-k_t/\mu)$,
as motivated by the observed transverse momentum distribution of direct 
pions in  hadroproduction (labeled "exp."). 
We consider also
a model where
$\phi(k_t \ge 1 {\rm GeV/c}) \propto {1 \over k_t^2}$ (labeled 
"reg.") in order to check the sensitivity of our results on the large 
transverse momentum tail of the nonperturbative wave function. 
Such a behavior (without the naively expected $\alpha_s(k_t^2)$) follows
directly from QCD \cite{privat}. At the same time, the region of
applicability of the pQCD prediction for the high momentum tail, which
we started at 1 GeV/c, is still subject of discussions.  We then 
vary $\mu$ in the range corresponding to average transverse momenta 
$\sqrt{\left< k^2_t\right>} \sim 300 \div 600$ MeV/c.  The results of these 
calculations are presented in Fig. 5.


\begin{figure}[htb]
\figa{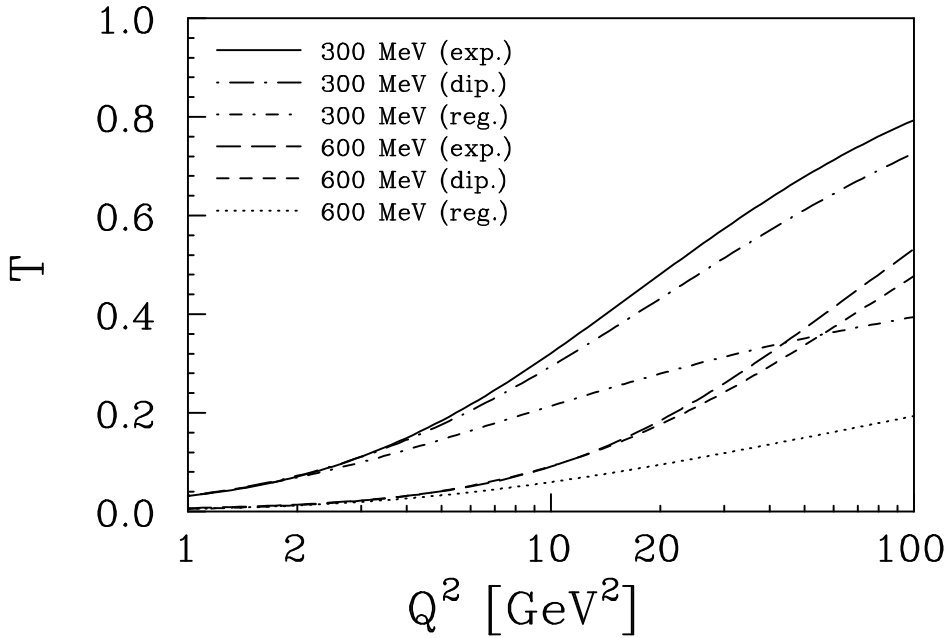}
\caption{Suppression factor, $T(Q^2)$ of Eq. (\protect\ref{tfactor}), in 
$d\sigma_{\gamma^*_LN \rightarrow VN}/dt|_{t=0}$ due to transverse quark Fermi 
motion for diffractively produced vector mesons built of light quarks.}
\end{figure}

        One can see that a very signicant suppression of the vector meson 
yield may be present in the $Q^2$ range covered so far by the HERA experiments,
and that the high momentum tail -- due to hard interactions between the 
constituents of the vector meson -- seems to be an important correction 
in the relevant range of $Q^2 \gtrsim 10$ GeV$^2$.  As can be seen from
Fig. 5, at large $Q^2$, $T$
is determined mostly by the high momentum tail of the $k^2_t$ 
distribution of the bare quarks within the diffractively produced vector meson. 
In mean field models of hadrons -- bag models, constituent quark models 
with an oscillator potential or 
models based on the method of dispersion sum rules -- this tail is included 
into the effective parameters, and thus these models give no clues for the 
value of this component, or its $k_t$ and $z$ dependence.  On the other hand,
as suggested by Fig. 5, a study of vector meson production over a wide range
of $Q^2$ would allow to obtain unique information on the $k_t$ distribution
in hadrons. Note, however, that to achieve a nonambigous interpretation
of quark Fermi motion effects, it is necessary to calculate also the 
contribution
of the $q \bar q g$ component in the light-cone wave functions of the 
$\gamma^*_L$ and the vector meson to the cross section of hard diffractive 
processes. This has not been done so far.

\subsection{Charmonium production}

        In the case of the photo- and electroproduction of heavy vector mesons,
the overlap integral is suppressed both due to the presence of a term 
proportional to $m_c^2$ and due to quark Fermi motion. 

        To investigate the role of Fermi motion for a realistic case, we 
consider the nonrelativistic model of charmonium. We keep terms proportional 
to ${k^2 \over m_c^2}$ in the wave function of the photon, as they enter
with a large numerical coefficient and because the integral over the 
charmonium wave function only slowly converges at large momenta. At the same 
time, we neglect relativistic effects in the charmonium wave function and 
set\footnote{This 
formula follows from the necessity to transform light-cone perturbation
theory diagrams into conventional, nonrelativistic diagrams describing
the charmonium bound state.}
$z\rightarrow {1\over 2}\left(1+{k_z\over 
\sqrt{k^2+m_c^2}}\right)$.
In this 
approximation, the phase volume in, for instance, Eq. (\ref{tfactor}) becomes
${dzd^2k_t \over z(1-z)} \rightarrow {2 \over \sqrt{k^2+m_c^2}}d^3k$, 
the energy
denominator transforms into ${m_c^2+k_t^2\over z(1-z)}\rightarrow 
4(m_c^2+k^2)$, 
and the charmonium's light-cone wave function $\psi_V(z,k_t)$ can be obtained
from its non-relativistic rest frame wave function $\psi_V(k)$ based on the
requirement that the $J/\Psi \rightarrow e^+e^-$ decay width is reproduced
correctly, which yields $\int dz\,d^2k_t\,\psi_V(z,k_t) = \int d^3k\,\psi_V(k)$.
In this limit, $\eta_V \approx 2.44$,
and the corresponding factor, $F_V$, is significantly smaller than unity:
\begin{equation}
F_V(Q^2)=\left({
\int d^3k\,\psi_V(k)
\left(-{1\over 16}\Delta_t\right)
\left[{Q^4 \over Q^2 + {k_t^2+m_c^2\over z(1-z)}} \right]
\over 
\int d^3k\,\psi_V(k)\,{k^2+m_c^2\over k_t^2+m_c^2}
}\right)^{\!\!2} \ .
\label{psifactor}
\end{equation}
Here, $\Delta_t$ is the two dimensional Laplace operator which acts on the
wave function of the photon.
For numerical estimates, we use two different realistic charmonium wave 
functions calculated from a QCD-motivated \cite{charm3} 
and a logarithmic 
potential \cite{charm2}, which both describe $\Gamma_{J/\Psi \rightarrow 
e^+e^-}$ reasonably well.  
We hereby restrict our consideration to potential models for which the mass
of the constituent $c$ quark is close to the mass of the bare current
$c$ quark, i.e. $m_c \approx 1.5$ GeV 
($m_c=1.48$ GeV for the QCD-motivated potential of Ref.
\cite{charm3}, and $m_c=1.5$ GeV for the logarithmic potential of Ref.
\cite{charm2}).
This is necessary to keep a minimal correspondence with the QCD
formulae for hard diffractive processes which are expressed through the
distribution of bare quarks \cite{Brod94}.
The results of these calculations are shown in 
Fig. 6 for diffractive electroproduction of $J/\Psi$ and $\Psi'$ mesons, 
respectively.  Fig. 6 shows that, in this model,
the applicability of the asymptotic formulae
requires rather large $Q^2$.


\begin{figure}[htp]
\figa{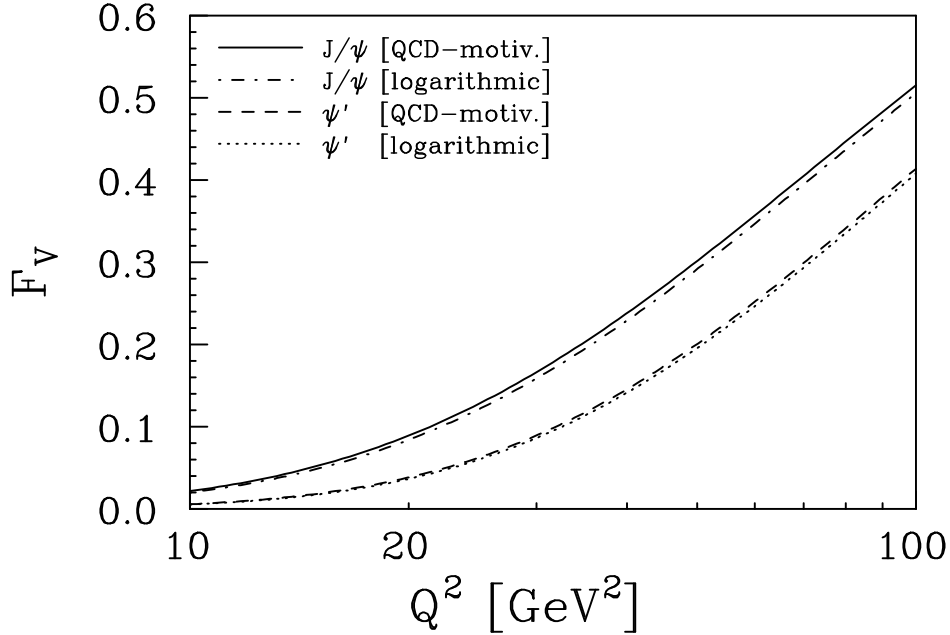}
\caption{Suppression factor, $F_V(Q^2)$ of Eq. (\protect\ref{psifactor}), in 
$d\sigma_{\gamma^*_LN \rightarrow VN}/dt|_{t=0}$ due to the finite 
quark mass and quark Fermi motion for diffractively produced $J/\Psi$ and 
$\Psi'$ mesons.}
\end{figure}

        Note that, in the evaluation of $\Gamma_{J/\Psi \rightarrow e^+e^-}$ 
and also in the process considered here, large radiative corrections,
$1-{16 \alpha_s(m_c^2) \over 3 \pi}$,
accompany the charmonium's non-relativistic wave function but {\em
not} the $q \bar q$ component of the
light-cone wave function, which underlines the limits
of applicability of non-relativistic charmonium potential models in that
context. 
Inclusion of these radiative corrections effectively 
corresponds to a renormalization of the nonrelativistic  
wave function by the factor $1-{8 \alpha_s(m_c^2)\over 3 \pi}$.  Since we
only employ the light-cone normalization of the wave function, this factor is 
implicitly included in all our calculations.
 Furthermore, the nonrelativistic charmonium model
does not include gluon emission at higher resolution, and it therefore does
not display the expected asymptotic behavior: $\int d^2k_t\,\psi_V(z,k_t)
\propto z(1-z)$.  In principle, this could be accounted for by means of
$b$-space evolution of the leading twist wave function as suggested by
Brodsky and Lepage \cite{BL}.

        In Ref. \cite{Ryskin}, pQCD was applied to the diffractive production 
of $J/\Psi$ mesons starting from photoproduction. The observed increase 
of the $J/\Psi$ photoproduction cross section with initial energy is in line 
with pQCD expectations \cite{Ryskin}. Moreover, the theoretical analysis of 
the transverse distances between the $c$ quarks within the $J/\Psi$ meson 
shows that they are close to that dominant in the wave function of the 
$\gamma^*$ in the deep inelastic structure functions. Thus, the necessary 
condition for the applicability of pQCD is the same in all cases. Specific to 
charmonium, however, is that these distances are close to the average 
distances between the quarks within the $J/\Psi$ meson, as was discussed in
the previous section. So, the process under consideration should be very 
sensitive to the exact form of the light-cone wave function of the $J/\Psi$ 
meson and to its color field.

        It was assumed in Ref. \cite{Ryskin} that the cross section of 
diffractive $J/\Psi$ electroproduction could be calculated employing the 
nonrelativistic constituent quark model for the $J/\Psi$ meson's wave 
function, both for transverse and longitudinal photon polarizations, and 
quark Fermi motion was neglected.  This corresponds to assuming that 
$R \equiv \eta_{J/\Psi} T_{J/\Psi}=2$.  In the limit where we can justify 
the application of pQCD ($M^2_{J/\Psi}\ll Q^2 $), our result coincides with 
that of Ref.~\cite{Ryskin} if we 
neglect Fermi motion and $m_c \ne M_{J/\Psi}/2$ effects, 
i.e., assume that $R=2$,
and if we discard 
the qualitative difference between the minimal $c \bar c$ Fock component in 
the $J/\Psi$'s wave function
and its wave function in nonrelativistic potential models,
which is actually not legitimate as was discussed in Sect. III.A.
Thus, in Ref.~\cite{Ryskin}, the predicted 
cross section is overestimated within the nonrelativistic quark model by a 
factor $T_V^{-1}(Q^2)$, where
\begin{equation}
T_V(Q^2)=\left({
\int d^3k\,\psi_V(k)
\left(-{1\over 16}\Delta_t\right)
\left[{(Q^2+4m_c^2)^2 \over Q^2 + {k_t^2+m_c^2\over z(1-z)}} \right]
\over 
\int d^3k\,\psi_V(k)\,{k^2+m_c^2\over k_t^2+m_c^2}
}\right)^{\!\!2} \ ,
\label{Rysfactor}
\end{equation}
which is presented in Fig. 7 for the charmonium wave functions employed
previously.


\begin{figure}[htb]
\figa{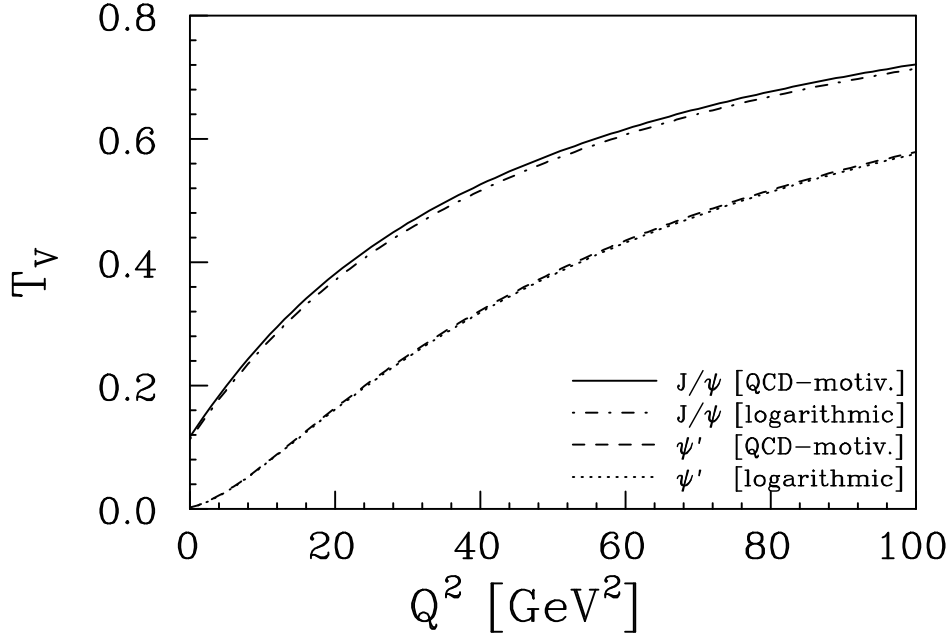}
\caption{Correction factor, $T_V(Q^2)$ of Eq. (\protect\ref{Rysfactor}), in 
$d\sigma_{\gamma^*_LN \rightarrow VN}/dt|_{t=0}$ for diffractive charmonium 
production due to quark Fermi motion.}
\end{figure}

        One can see from Fig. 7, that the effects of Fermi motion strongly 
suppress the contribution of the cross section, especially for 
$Q^2 \approx 0$ where most of the experimental data were obtained. 
If we evaluate the suppression factor $T_V$ of Eq. (\ref{Rysfactor})
at the photoproduction point with charmonium wave functions calculated 
from the two nonrelativistic potentials of Refs. \cite{charm2,charm3},
we find a value of $T_V(Q^2=0) \approx 1/9$. This is 
significantly smaller than the corresponding value ($\sim 0.5$)
given recently by Ryskin et al. \cite{Ryskin2}.\footnote{In our opinion, 
the reason for this discrepancy is 
threefold: Firstly, the authors of Ref. \cite{Ryskin2} use the approximate
Fermi motion correction which we gave in Ref. \cite{hepph}, secondly they
neglect the longitudinal relative motion 
of the quarks in the $J/\Psi$ mesons -- by setting $z={1\over 2}$ -- which, 
in turn, transforms the three dimensional integral over $d^3k$ in Eq. 
(\ref{Rysfactor}) into a two dimensional integral over $d^2k_t$  -- see Eq. 
(23) in Ref. \cite{Ryskin2} -- and, thirdly, a Gaussian form for the wave 
functions was assumed in Ref. \cite{Ryskin2}, whereas the realistic potentials 
of Refs. \cite{charm1,charm2,charm3} all yield wave functions that 
fall off significantly more slowly.  All three
approximations diminish the role of large quark momenta and result in 
an essential reduction of the suppression factor. For instance,
our $T_V(Q^2=0)$ would rise from 0.11 to 0.4 if we were to 
use our approximate formula of Ref. \cite{hepph} and change the
integration over $d^3k$ in Eq. (\ref{Rysfactor}) to an integration
over $d^2k_t$, and it would rise further to 0.5 if we would replace
our realistic charmonium wave functions with Gaussians. Note that the
latter value coincides with that given in Ref. \cite{Ryskin2}.}

$T_V(0)$ has been evaluated also by decomposition over powers of
${k^2\over m_c^2}$: $T_V(0)=1-{20\over 3} \langle{k^2\over m_c^2}\rangle$.
The difference between the actual number of $20/3$ and the $4$ obtained in
Ref. \cite{Ryskin2} is due to account of all terms arising from the
application of the $\Delta_t$ operator.
By definition, $\langle k^2\rangle ={\int k^2 \psi_V(k) d^3k \over 
\int \psi_V(k) d^3 k}$. In Ref.
\cite{Ryskin2}, Gaussian type wave functions for the $J/\psi$ were
used to estimate that $\langle k^2/m_c^2 \rangle \approx 1/8$ (Note that
for a pure Coulomb potential,
which, when masked with a running coupling constant,
corresponds to the limit $m_c \rightarrow \infty$ in QCD,
$\langle k^2\rangle=\infty$ due to a
linear divergency at large $k$.).  So, within this approximation, 
$T_V(0) \approx 1 / (1+\alpha_s)^6$.

The suppression factor $T_V(Q^2)$ has
contributions from leading and non-leading twist. The respective leading 
twist expression follows from Eq. (\ref{tfactor}) 
by setting, in the wave function
of the photon, $k_t$ to zero after differentiation:
\begin{equation}
T_V^{LT}(Q^2)=\left({
\int {dz\,d^2k_t\over z(1-z)}\,\psi_V(z,k_t)
\left[{(Q^2+4m_c^2)\over Q^2 + {m_c^2\over z(1-z)}} \right]^2
\over 
\int {dz\,d^2k_t\over z(1-z)}\,\psi_V(z,k_t)}\right)^{\!\!2} \ .
\label{leadingtwist}
\end{equation}
Additional (leading twist) terms containing $\ln m_c$  will 
be accounted for in a separate publication.
Our analysis shows that, in QCD, the higher twist corrections are 
smaller, which, in turn, justifies our 
discussion of those Fermi 
motion effects in what is otherwise strictly a leading
twist analysis.

        Furthermore, the results of calculations of $J/\Psi$ and $\Psi'$ 
photoproduction cross sections are rather sensitive to the value used for 
the mass of the $c$ quark, and, for instance, in Ref. \cite{Ryskin} the 
assumption was made that $m_c=M_{J/\Psi}/2$. If we analyze the dependence 
of the $J/\Psi$ photoproduction cross section on the difference between
$M_{J/\Psi}$ and $2m_c$, we find a correction factor,
\begin{equation}
R_{M_{J/\Psi} \ne 2m_c}
=\left({M_{J/\Psi}\over 2m_c} \right)^{\!6}
\left({
\int d^3k\,{\psi_V(k)\over (1 + k^2/m^2_c)^2}
         \,{k^2+m_c^2\over k_t^2+m_c^2}
\over  
\int d^3k\,{\psi_V(k)\over (1 + 4k^2/M^2_{J/\Psi})^2}
         \,{k^2+m_c^2\over k_t^2+m_c^2}
}\right)^{\!2} \ ,
\label{masfactor}
\end{equation}
which, using the values for $m_c$ and the respective charmonium wave 
functions from Refs. \cite{charm2,charm3}, yields
$R_{M_{J/\Psi} \ne 2m_c} \approx 1.2$. This, together with our
$T_V(Q^2=0)$, gives a final prediction for the suppression of 
$J/\Psi$ photoproduction of\footnote{In Ref. 
 \cite{Ryskin2}, the correction factor
 of Eq. (\ref{masfactor}) was approximated by $(M_{J/\Psi}/2m_c)^8$,
 and the latter was evaluated with a small $m_c=1.43$ GeV,
 as estimated by Jung et al. \cite{jung}, which, in turn, yields 
 a very strong enhancement of $R_{M_{J/\Psi} \ne 2m_c} \approx 2$.
 This estimate is at variance with the direct calculation using 
 Eq. (\ref{masfactor}).
 This, together with the underestimated Fermi motion correction, leads the 
 authors of Ref. \cite{Ryskin2} to the conclusion that there is no net 
 suppression for $J/\Psi$ photoproduction.} $\approx 1/8$.

\subsection{$Q^2$ rescaling}

        In Sect. III.A, we determined the relevant transverse distances
between the quark and the antiquark in the hard blobe of Fig. 1, and we
found that they depend on the process under consideration, i.e. there
is a substantial difference between $b_{\sigma_L}(Q^2)$
and $b_{\gamma^*_L\rightarrow V}(Q^2)$.

        In detail, we used the impact parameter representation of the 
longitudinal structure function, $\sigma_L(x,Q^2)$ of Eq. (\ref{sigmabb}), 
to define a relation between the transverse size of the $q\bar q$ pair, $b$, 
and the virtuality of the process, $Q^2$.  This relation, 
\begin{equation}
b = b_{\sigma_L}(Q^2) \approx {9.2 \over Q^2} \ ,
\label{scal1}
\end{equation}
at $x=10^{-3}$, yields the scale effective in the gluon densities
which enter, for instance, 
in the elementary color-dipole cross section, $\sigma(b^2)$
of Eq. (\ref{eq7}). It is designed
in such a way that, if we were to evaluate the integrand of Eq. (\ref{sigmabb})
at this average $b=b_{\sigma_L}(Q^2)$ instead of carrying out the integral, we 
would just recover the normalization condition: $\sigma_L(x,Q^2) \propto 
\alpha_s(Q^2)\,xG_N(x,Q^2)$.  Hence, the virtuality that corresponds to the
dominant transverse distances and the virtuality of the process are identical,
at least for the case of the longitudinal photoabsorption cross section.

        However, this is not the case for the electroproduction amplitude,
${\cal A}_{\gamma^*_LN\rightarrow VN}$ of Eq. (\ref{brep}), due to the
significant difference between the transverse distances it is dominated
by and $b_{\sigma_L}(Q^2)$.
This indicates that substantial next-to-leading order
corrections should be present in diffractive vector meson photo-
and electroproduction which, in turn, should lead to a change of the scale
effective in the gluon densities, i.e., the virtualities that enter in the
argument of $\alpha_s(Q^2)\,xG_N(x,Q^2)$ in our basic equation (\ref{eq5a}).

        Following our logic of approximating the integrals over the transverse
sizes of the $q\bar q$ pair effective in the hard blobe of Fig. 1 by the 
contribution from an appropriately defined average transverse distance,
a natural guess is to rescale the virtuality $Q^2$ in proportion of these
average transverse sizes, 
\begin{equation}
Q_{eff}^2 \approx Q^2\,
{b^2_{\sigma_L}(Q^2) \over b^2_{\gamma^*_L\rightarrow\rho}(Q^2)} \ ,
\label{scal2}
\end{equation}
which, for $\rho$ production, yields significantly smaller virtualities,
i.e., $Q^2_{eff} < Q^2$.
This results in a strong decrease of the quantity 
$[\alpha_s(Q^2)\,xG_N(x,Q^2)]^2$, and
therefore also of the $\rho$-meson electroproduction cross section,
as can be seen in Fig. 8 where we show $[\alpha_s(Q^2)\,xG_N(x,Q^2)]^2$
with and without the NLO rescaling.\footnote{Eq. (\ref{scal2}) assumes that 
$G_N(x,Q^2)$ is defined such that Eq. (\ref{sigmat}) for $\sigma_L$ is valid 
in the NLO approximation.  This question requires further investigation. 
However, the corresponding uncertainty is numerically unimportant for
the comparison of production rates of different vector mesons.}


\begin{figure}[htp]
\figf{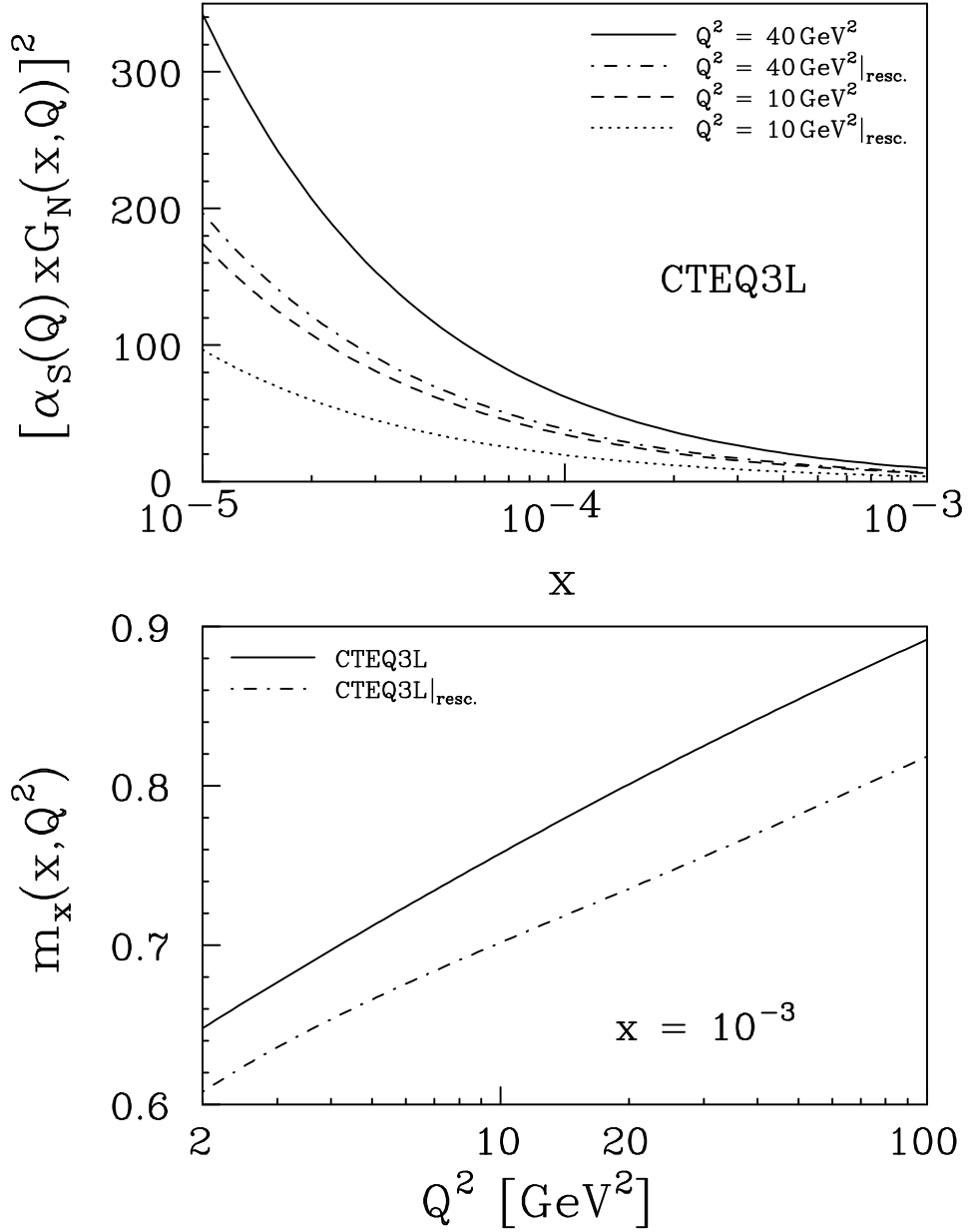}
\caption{The square of the gluon density, $[\alpha_s(Q^2)\,xG_N(x,Q^2)]^2$,
and its energy dependence, $m_x(Q^2)$, defined by 
$[\alpha_s(Q^2)\,xG_N(x,Q^2)]^2 \propto x^{-m_x(Q^2)}$
calculated for the CTEQ3L parton distribution function with and without the
NLO ($\rho$-meson electroproduction) rescaling introduced in Eq. 
(\protect\ref{scal2}).}
\end{figure}

        In Fig. 8, we also plot the exponent of the energy
dependence of $[\alpha_s(Q^2)\,xG_N(x,Q^2)]^2$ (and hence of the 
predicted cross sections), $m_x(Q^2)$, which is defined by means 
of $[\alpha_s(Q^2)\,xG_N(x,Q^2)]^2 \propto x^{-m_x(Q^2)}$.  We observe
that the rescaling not only significantly reduces the predicted
cross sections, but it also results in a slight reduction of its steep
energy dependence.

        On the other hand, the relevant transverse distances for
$J/\Psi$ photoproduction, $b_{J/\Psi}(Q^2=0)$, are relatively small,
$b_{J/\Psi}(Q^2=0) \approx 0.25$ fm, which, using the logic outlined in the
above, results in substantially higher effective virtualities,
$\overline{Q}^2 \approx {9.2 \over b_{J/\Psi}^2} \approx 5.1$ GeV$^2$, for
$J/\Psi$ photoproduction than those suggested, for instance, in Ref.
\cite{Ryskin2} -- $\overline{Q}^2 = 2.4$ GeV$^2$. This will be discussed
in more detail in Sect. V.C.

        In Fig. 9, we finally show the $Q^2_{eff}$ obtained in this manner
for the cases of $\rho$-meson electroproduction and $J/\Psi$ photo- and
electroproduction.  Note that the relationship between $Q^2$ and $Q^2_{eff}$
is almost independent of $x$, and that the $Q^2_{eff}$ for $J/\Psi$ production
differs significantly from the estimate of Ref. \cite{Ryskin2},
$\overline{Q}^2 = {Q^2 + M_{J/\Psi}^2 \over 4}$, which is shown through 
the dotted line in Fig. 9b.


\begin{figure}[htp]
\figf{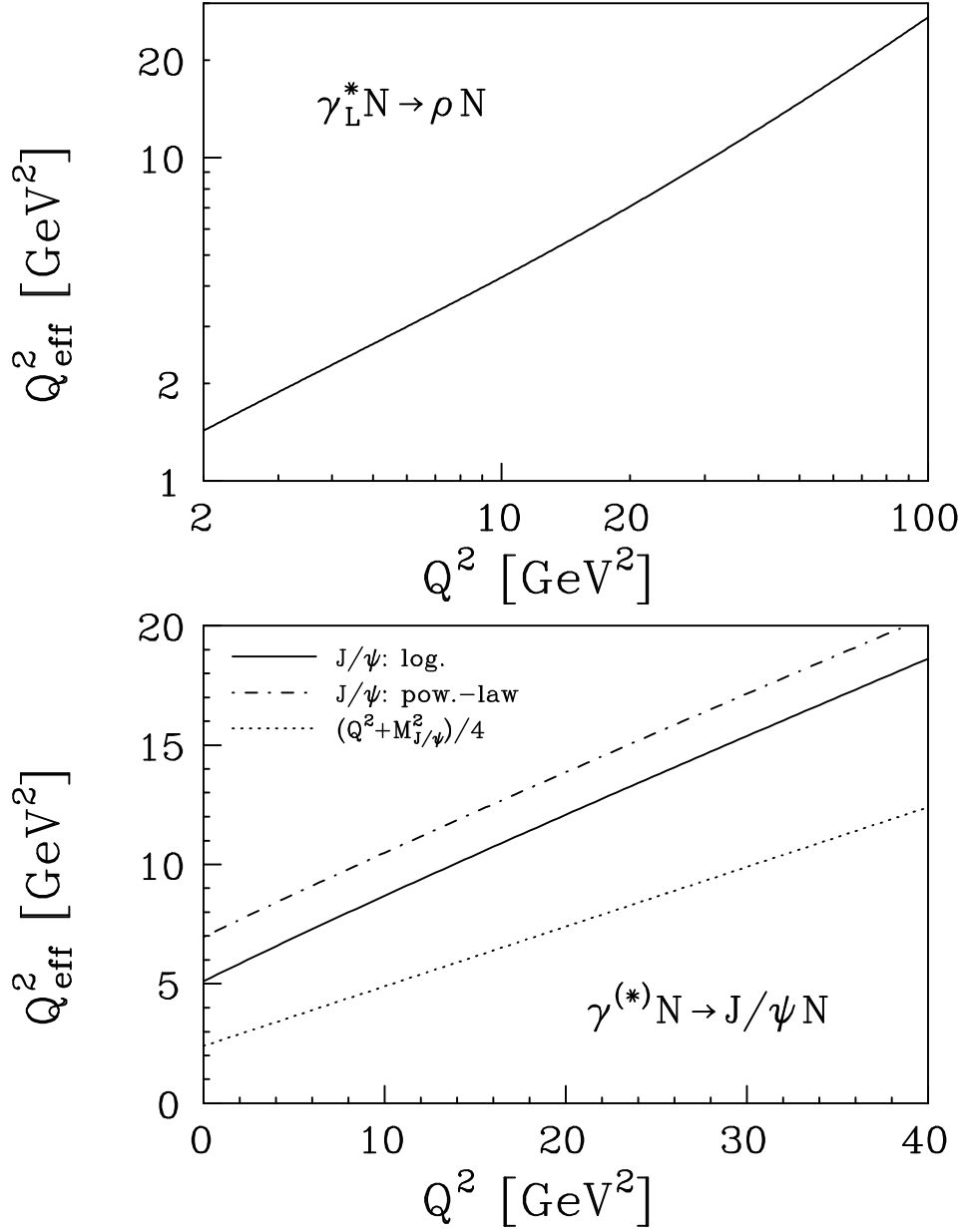}
\caption{$Q^2_{eff}$ obtained via the NLO $Q^2$ rescaling introduced in
Eq. (\protect\ref{scal2}) for $\rho$-meson 
electroproduction and $J/\Psi$ photo- and electroproduction.
For the latter, we also show the estimate of Ref. \protect\cite{Ryskin2}
(the dotted line).}
\end{figure}
 
\section{Comparison with data}
\subsection{Main features of the data}

        The HERA data~\cite{ZEUS}, when combined with the NMC data \cite{NMC1} 
on diffractive electroproduction of $\rho$-mesons, are consistent with
several key predictions of Eq.~(\ref{eq5a}): 

(i) a fast increase with energy of the cross section for electroproduction of 
vector mesons, which is approximately proportional to $x^{-0.6}$ for $Q^2=10$ 
GeV$^2$. Note that this fast increase with decreasing $x$ is not expected 
within the non-perturbative two-gluon-exchange model of Donnachie and Landshoff
\cite{DL1}, which predicts a cross section rising as $\sim x^{-0.16}$ at $t=0$,
and an even weaker increase of the cross section when integrated over $t$ due 
to a substantial increase of the slope of the elastic cross section expected 
in soft Pomeron models: $\sigma_{el} \propto {1 \over x^{0.16}} {B_0 \over B_0 
+ 2 \alpha' \ln{x_0 \over x}}$. Here, $B_0$ is the slope of the elastic cross 
section for $x=x_0$.   

(ii) the dominance of the longitudinal polarization, ${\sigma_L \over 
\sigma_T}\propto Q^2$ \cite{Brod94,DL1}.
Data do confirm the dominance of $\sigma_L$, though 
they are not accurate enough to study details of the $Q^2$ dependence. 
Predictions for the $x$-dependence of this ratio sensitively depend on the 
interplay of perturbative and nonperturbative contributions to $\sigma_T$.
In the leading $\alpha_s \ln{Q^2\over\lambda_{QCD}^2}$ approximation,
it is easy to generalize the QCD evolution equation to transversely polarized
photons as a projectile \cite{Halina}. As in the case of the regular 
structure functions, the appropriate question will be on the region of 
integration over fractions of momenta appropriate to the hard blob. Since the 
hard kernel in the evolution equation increases faster at small $x$ than
the soft contribution, one may expect an onset of hard physics for
transversely polarized photon projectiles as well, though at smaller $x$ and at 
higher $Q^2$. Note that at sufficiently large $Q^2$, the contribution of 
nonperturbative QCD to diffractive vector meson production is suppressed in 
addition by a Sudakov type form factor.  At sufficiently large $Q^2$, 
the process should be discussed in terms of the Bethe-Salpether wave function 
for the transition of a $\rho$-meson into two highly virtual quarks. 
Such a mathematical object exists in pQCD only.  Thus, there exists a 
possibility to check quark counting rules in their original form.
This may manifest itself in the $x$-dependence of the ratio ${\sigma_L \over 
\sigma_T}$ for fixed $Q^2$.  At intermediate $Q^2 \sim 10$ GeV$^2$, where hard 
physics already dominates in $\sigma_L$, $\sigma_T$ may still be determined 
by soft, nonperturbative contributions.  For these $Q^2$, the ratio should
increase with decreasing $x$ as $x^2G_N^2(x,Q^2)$.  At high $Q^2$,
where hard physics dominates for both $\sigma_L$ and $\sigma_T$, the
ratio would not depend on $x$. We are currently investigating these issues
in detail \cite{progress}.  The $x$ dependence of ${\sigma_L \over \sigma_T}$ 
was also discussed in Ref. \cite{nemch2}. The authors employed nonrelativistic 
oscillator wave functions for the diffractively produced vector mesons, and
the pQCD color-dipole picture was applied for both polarizations and 
also for large transverse distances. For both cases, the formula for
the elementary color-dipole cross section, $\sigma(b^2)$ of Eq. (\ref{eq7})
which was deduced in \cite{BBFS93,FMS93}, is clearly not applicable,
as was discussed in detail in Sect. III.

(iii) the absolute magnitude of the cross section is reasonably well
explained,
though uncertainites in the knowledge of $xG_N(x,Q^2)$ are rather large 
to make a comparison quantitative. This will be discussed in 
detail in the next subsection.

(iv) the $Q^2$ dependence observed by the ZEUS collaboration: $\sigma_L 
\propto Q^{-n}$ with $n= 4.2 \pm 0.8$ \cite{ZEUS}.  This seems to reflect 
the $Q^2$ evolution of the parton distributions as well as effects of quark 
Fermi motion within the vector mesons, as will be outlined in subsection V.D.

\subsection{Absolute magnitude of the $\rho$-electroproduction cross section}

Currently, absolute cross sections for exclusive $\rho$-meson production 
are available from NMC \cite{NMC1} and from ZEUS \cite{ZEUS}.

We restrict our analysis here
to $Q^2 \ge 6$ GeV$^2$. In the case of NMC, the most 
relevant data are those obtained using a deuteron target as
no data were taken with hydrogen. The measurements do not 
separate the elastic deuteron final state and deuteron break up. So, to extract 
the cross section for $\rho$-production off the nucleon from the deuteron
data, we use the closure approximation to sum over the final states of the two
nucleons, which yields \cite{bauer}
\begin{equation}
\frac{d\sigma_{\gamma^* D \rightarrow \rho D}}{dt} =
2(1+F_D(4t)) \frac{d\sigma_{\gamma^* N \rightarrow \rho N}}{dt} \ ,
\label{equa1} 
\end{equation}
where $F_D(t)$ is the deuteron's electromagnetic form factor. In the above,
we neglected shadowing, Glauber rescattering effects (which are suppressed
due to color transparency expected in this kinematics) and the real part of
the elementary amplitude. $F_D(t)$ can be roughly parameterized as
$F_D(t) \sim e^{B_Dt}$ with $B_D \approx 14$ GeV$^{-2}$ \cite{bauer}. This 
yields 
\begin{equation}
\sigma_{\gamma^* N \rightarrow \rho N} =
\frac{\sigma_{\gamma^* D \rightarrow \rho D}}{2}
\left(1-\frac{B}{4B_D}\right) \ ,
\label{equa2} 
\end{equation}
where $B=4.6 \pm 0.8$ GeV$^{-2}$ \cite{NMC1} is the slope of the 
$\rho$-production cross section off the nucleon. With this, the total cross 
section of $\rho$-production on the nucleon, $\sigma_{\gamma^* N \rightarrow 
\rho N}$, can be calculated from the NMC's deuteron data. Using their 
determination of the virtual photon's polarization, $\langle\epsilon\rangle$, 
and the measurement of $R=\sigma_{\gamma^* N \rightarrow \rho N}^L/
\sigma_{\gamma^* N \rightarrow \rho N}^T =2.0 \pm 0.3$ at $\langle Q^2\rangle 
= 6$ GeV$^2$, we can furthermore separately extract the elementary longitudinal 
cross section, $\sigma_{\gamma^* N \rightarrow \rho N}^L$. In the same manner, 
$\sigma_{\gamma^* N \rightarrow \rho N}^L$ can be obtained from the ZEUS data 
where $R=1.5 \pm 0.6$ (statistical errors only) at $\langle Q^2 \rangle = 11$ 
GeV$^2$.


\begin{figure}[htp]
\figd{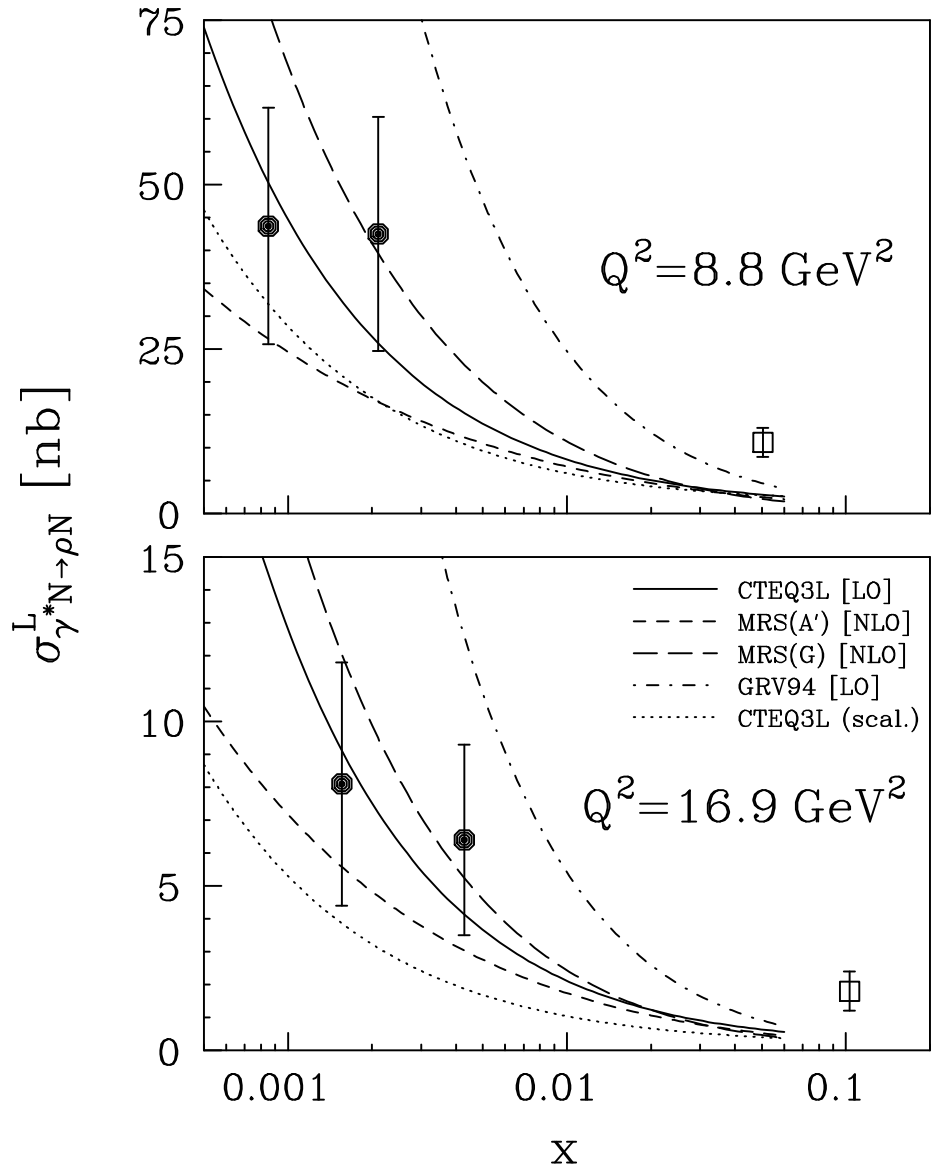}
\caption{The total longitudinal cross section, $\sigma_{\gamma^* N \rightarrow
\rho N}^L$, calculated from Eq. (\protect\ref{eq5a}) for several recent
parametrizations of the gluon density in comparison with experimental data
from ZEUS \protect\cite{ZEUS} (full circles) and NMC \protect\cite{NMC1}
(squares).}
\end{figure}

Results of such calculations are presented in Fig. 10. There, we also show 
theoretical predictions of Eq.~(\ref{eq5a}) employing several recent 
parametrizations of the gluon 
density\footnote{Note that, within the leading $\ln{Q^2\over\Lambda_{QCD}^2}$ 
approximation to QCD, the use of NLO fits to the nucleon's gluon density
is not fully consistent.}
\cite{MRS,CTEQ,GRV} and typical 
parameters 
for the $\rho$-meson wave functions as considered in section IV. We set 
$\eta_V=3$ and parametrize the dependence of the differential cross section 
on the momentum transfer in exponential form with\footnote{A natural estimate 
for $B$ is $B={\langle r_g^2 \rangle\over 3}+B_{\gamma^*_L\rho}(Q^2)$ where 
$\sqrt{\langle r_g^2 \rangle}$ is the mean radius of the gluon distribution
in a nucleon which is usually taken in realistic models of the nucleon
as 0.6 fm. The calculation of Sect. III.A gives $B_{\gamma^*_L\rho}(Q^2)
\approx 0.7$ GeV$^{-2}$. Putting all numbers together we obatin
$B \approx 3.8$ GeV$^{-2}$.}
$B \approx 4$ GeV$^{-2}$.
Note that a change of $T(Q^2)$ in the range depicted in Fig. 5 introduces 
an extra scale uncertainty of $0.5 \div 2$.

We discussed in Sect. IV.C that next-to-leading order corrections should 
lead to a decrease of the scale effective in the gluon densities, i.e., the 
virtualities that enter in the argument of $\alpha_s(Q^2)G_N(x,Q^2)$, due
to the fact that $b_{\gamma^*_L\rightarrow\rho}(Q^2)$ is
substantially larger than $b_{\sigma_L}(Q^2)$.
This results in a decrease of the $\rho$ production cross section as
illustrated in Fig. 10 for the CTEQ3L parametrization through the dotted lines.

One can see from the figure that we substantially underestimate the cross 
section at lower energies which, in turn, correspond to rather large $x$: 
for example $\langle x \rangle = 0.103$ for the NMC point at $Q^2=16.9$ GeV$^2$.
This kinematics is well beyond the $x$-range specified in Eq. (\ref{cohlen}),
and the frozen approximation, which we use, is inapplicable. 
Qualitatively, one can expect that, in this situation, quantum diffusion of the 
$q \bar q$ pair would lead to an increase of its interaction cross section, 
and hence to agreement with the observed cross section.

\subsection{Absolute magnitude of the $J/\Psi$ photoproduction cross section}

In a recent work by Ryskin et al. \cite{Ryskin2}, $J/\Psi$ photoproduction
data were compared with the corresponding pQCD predictions.  For the latter, 
the nonrelativistic constituent quark model for the $J/\Psi$'s wave function
was employed, and Fermi motion and $M_{J/\Psi} \ne 2m_c$ corrections were
estimated to leave the cross section unaltered.  On the other hand, we
argued in Sect. IV.B that there is a strong reduction ($\sim 0.13$) 
of the naive pQCD prediction for $J/\Psi$ photoproduction due to those 
presymptotic effects.  Here, we thus compare our predictions with the data,
while either accounting for the preasymptotic corrections, i.e. the factor 
$T_V(Q^2)$ of Eq. (\ref{Rysfactor}) and the $M_{J/\Psi} \ne 2m_c$
correction of Eq. (\ref{masfactor}), or while leaving them out.

Setting, $\eta_V = 2$ and 
$Q^2 \rightarrow \overline{Q}^2 = {Q^2 + 4m_c^2\over 4}$
as suggested in Ref. \cite{Ryskin2},
our Eq. (\ref{eq5a}) yields
\begin{equation}
\left. {d\sigma^L_{\gamma^{(*)}N\rightarrow J/\Psi N}\over dt}\right|_{t=0} =
{\pi^3\Gamma_{J/\Psi \rightarrow e^{+}e^-} M_{J/\Psi}\alpha_s^2(\overline{Q})
T_V(Q^2) \over 12 \alpha_{EM} \overline{Q}^6}
\Bigg|\left(1 + i{\pi\over2} {d \over d\ln x}\right) xG_N(x,\overline{Q})
\Bigg|^2 \ ,
\label{eq5ch}
\end{equation}
which agrees with the result given in \cite{Ryskin,Ryskin2} if we neglect
the Fermi motion and $M_{J/\Psi} \ne 2m_c$ corrections, i.e., 
if we set $T_V(Q^2)$
of Eq. (\ref{Rysfactor}) to 1 and replace $m_c$ by $M_{J/\Psi}/2$.  
On the other hand, we can also use the results of our analysis on the
relevant transverse distances of Sect. III.A (see Fig. 4)
to estimate the scale, $\overline{Q}$, effective in 
$\alpha_s(\overline{Q})xG_N(x,\overline{Q})$ in Eq. (\ref{eq5ch}).  
As outlined in Sect. IV.C, we find for $J/\Psi$
photoproduction $\overline{Q}^2 \approx 5.1$ GeV$^2$ which is substantially
higher than the estimate of Ref. \cite{Ryskin2} -- $\overline{Q}^2 = 2.4$ 
GeV$^2$ -- and which leads to an increase of the respective cross sections
by a factor of about 1.5 for the logarithmic potential of Ref. \cite{charm2}.
Taking the experimental value for the slope parameter of
$B_{J/\Psi} \approx 4.5$ GeV$^{-2}$, the cross section for diffractive
$J/\Psi$ photoproduction is calculated from Eq. (\ref{eq5ch}), and
corresponding results are shown in Fig. 11 for various 
recent parametrizations of the gluon density.


\begin{figure}[htp]
\fige{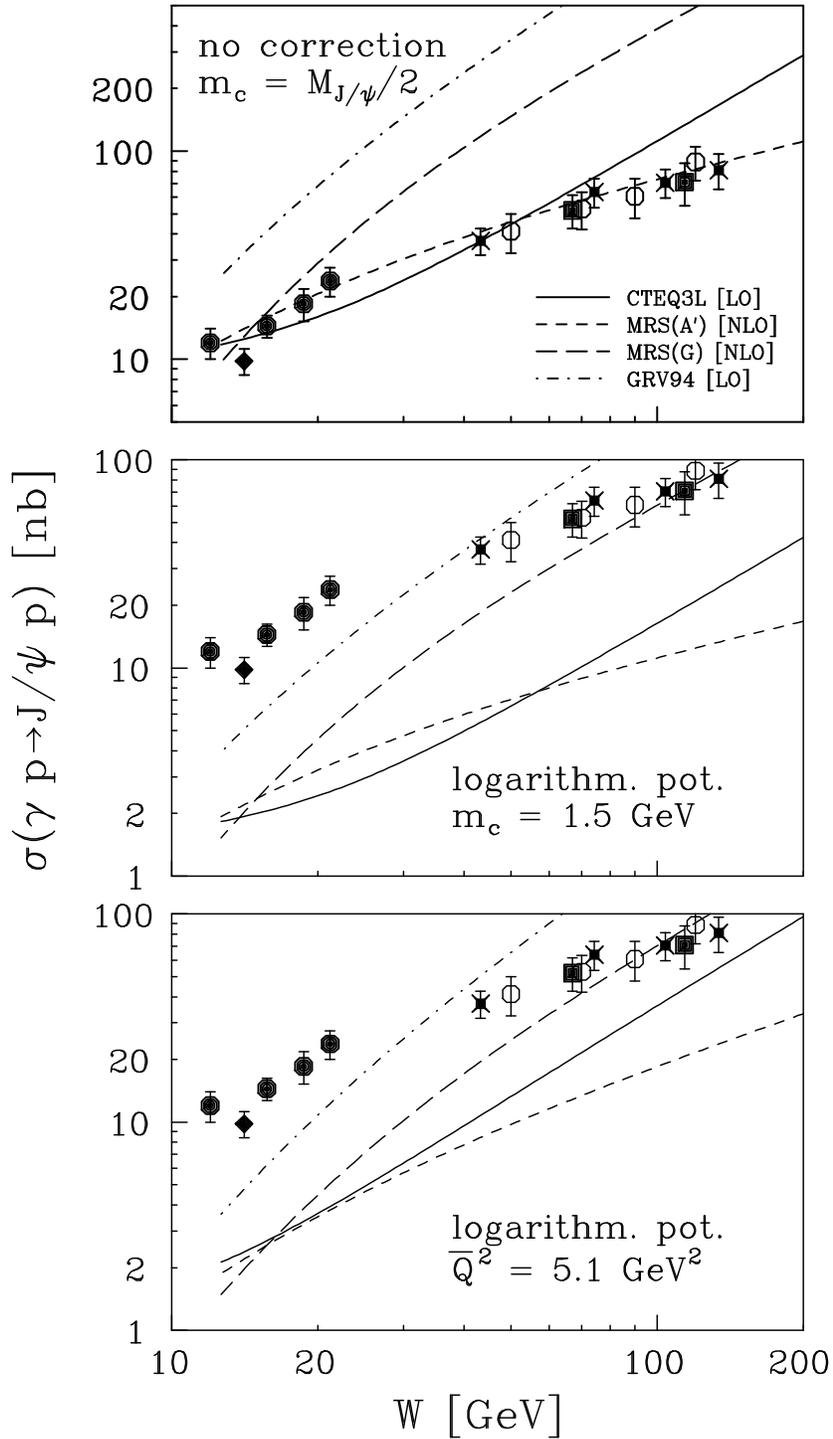}
\caption{The diffractive $J/\Psi$ photoproduction cross section 
compared with experimental
data from E401 \protect\cite{e401} (full circles), E516 \protect\cite{e516} 
(diamonds), ZEUS '93 \protect\cite{ZEUS1} (squares), and preliminary
ZEUS '94 (circles) and H1 (stars) data \protect\cite{prelim}. 
(a): no corrections ($\overline{Q} = M_{J/\Psi}/2$),
(b): with the corrections of Sect. IV.B ($\overline{Q} = m_c$)
(c): ditto (but with $Q^2$ rescaling: $\alpha_s xG_N$ evaluated at 
 $\overline{Q}$ based on $b_{J/\Psi}$).}
\end{figure}

        Thus, if Fermi motion, the $m_c \ne M_{J/\Psi}/2$ correction 
and $Q^2$ rescaling are accounted for, the predictions of the pQCD model 
and the photoproduction data agree, at large $W$, within the 
uncertainties in the existing gluon distributions
for the logarithmic potential model \cite{charm2}, which gives 
a smaller suppression as it has a $m_c$ close to the current quark mass
(cf. discussion in Sect. IV.B).

It is instructive, also, to consider the ratio of the $J/\Psi$ photoproduction 
and the longitudinal $\rho$-meson electroproduction cross sections,
$\sigma(\gamma\,p\!\rightarrow\!J/\Psi\,p)/\sigma(\gamma^*_L\,p\!\rightarrow
\!\rho\,p)$, for $Q^2=8.8$ GeV$^2$ and at the same $x$.  
We find that the various
gluon distributions lead to very similar ratios since the gluon densities enter
at about the same virtualities.  In detail, we find
$\left.\sigma(\gamma\,p\!\rightarrow\!J/\Psi\,p)/\sigma(\gamma^*_L\,p\!
\rightarrow\!\rho\,p) \right|_{theo} 
\approx 0.9$ with the $\overline{Q}$
rescaling and 0.6 without 
both at $x=2.1\cdot 10^{-3}$ 
and $x=8.5\cdot 10^{-4}$ 
with a theoretical error $\approx \pm 0.4$ due to uncertainties of 
the estimate of the suppression factor $T_V(Q^2)$.
The experimental data yield $1.2 \pm 0.6$ at
$x=2.1\cdot 10^{-3}$
and $1.6 \pm 0.7$ at $x=8.5\cdot 10^{-4}$, and thus the theoretical prediction 
for this ratio is no more than 
$1\sigma$ below the value we extract from the data.

Note, also, that, in a truely relativistic description, for photoproduction --
where only transversely polarized photons contribute -- the suppression factor 
of Eq. (\ref{Rysfactor}) would contain an extra factor $\propto 
{1 \over 4z(1-z)}$, 
which, in turn, would reduce the suppression by about 20\%.
At the same time, the analysis of the essential interquark distances (in Sect.
III.A) seems to indicate that the coupling of the two gluons 
to the charmonium meson is more complicated than assumed in the 
nonrelativistic description. In particular, the interaction of 
the exchanged gluons with the interquark potential is not suppressed.
Only in the approximation where the mass of $c$ quark tends 
to infinity, this effect would be small ($\sim {l_t^2 \over (m_c\alpha_s)^2}$)
where $l_t$ is the transverse momentum of the exchanged gluon.

One should also notice that in the fixed target experiments \cite{e401,e516},
the recoil proton was not detected and the corresponding cross sections could
thus have been overestimated.  Also, it was demonstrated
by Jung et al. \cite{jung} that, in the kinematics of these experiments,
the photon-gluon fusion
mechanism could yield up to 50\% of the cross section.
Besides, the coherence condition of Eq. (\ref{cohlen}) is violated for 
$J/\Psi$ photoproduction for $W \le 12$ GeV.

Also, the qualitative
agreement of our predictions for the $J/\Psi$ photoproduction
cross section with the data 
hints that the same approach could be used to calculate the
total $J/\Psi$-N interaction cross section.
The inspection of Fig. 3 
-- at $b \approx 0.40$ fm --
shows that such a model approach predicts a rather
fast increase of this cross section with energy (in this case 
$x=M^2_{J/\Psi}/s$) which may appear important for the interpretation
of $AA$ collisions. 
Note that the transverse distances relevant for $J/\Psi$ 
scattering are larger than those for
$J/\Psi$ photoproduction where small $b$'s are enhanced
due to the presence of the photon's wave function in
the integral.
 
\subsection{$Q^2$ dependence of $\rho$-meson electroproduction}
 

\begin{figure}[htp]
\figd{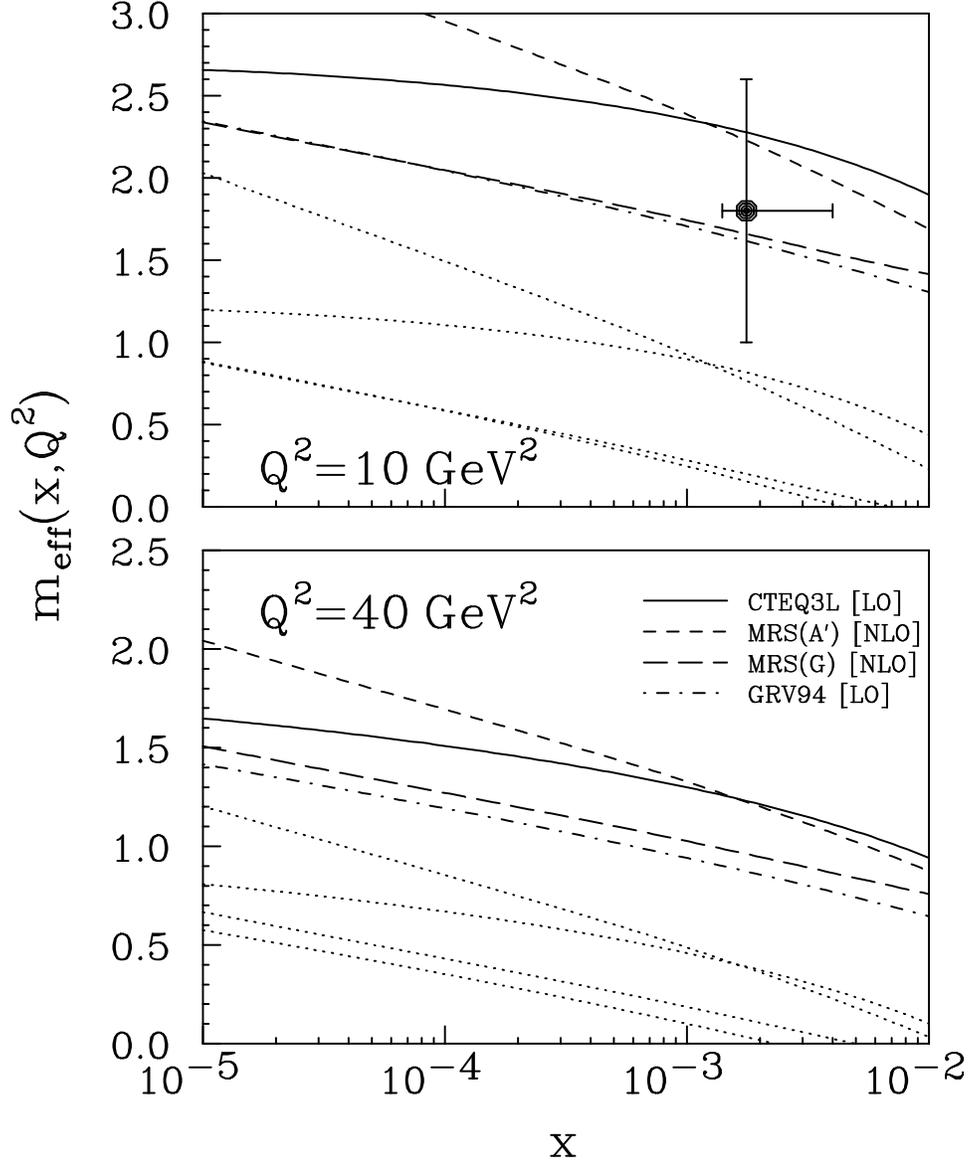}
\caption{Effective power of the $Q^2$ dependence of $(\alpha_s(Q) \,
x G_N(x,Q))^2$ as defined in Eq. (\protect\ref{neff}) (dotted lines) and 
including preasymptotic effects from quark Fermi motion together with the
ZEUS data point \protect\cite{ZEUS}.}
\end{figure}
 
        Let us now analyze the predictions of Eq.~(\ref{eq5a})
for the $Q^2$ dependence of the vector meson yield. One may conclude from   
inspection of this equation that it leads to a $1/Q^6$ dependence of the
cross section. There are, however, two QCD effects which substantially modify 
this naive expectation. At large $Q^2$, where higher order corrections 
can be neglected, Eq.~(\ref{eq5a}) predicts a $Q^2$ dependence of
the cross section which is substantially slower than $1/Q^6$ because
the gluon densities fastly increase with $Q^2$ at small $x$.
Numerically, the factor $(\alpha_s(Q) \, xG_N(x,Q))^2$ in Eq.~(\ref{eq5a}) 
is $\propto Q^n$ with $n \sim 1$. To illustrate this point, we define the 
effective power of the $Q^2$ dependence of $[\alpha_s(Q) \, xG_N(x,Q)]^2$ as
\begin{equation} 
m_{eff}(x,Q^2) \equiv 2\,{\partial \ln \left(\alpha_s(Q) x G_N(x,Q) \right) 
 \over \partial\ln Q} \ .
\label{neff} 
\end{equation}

Results for $m_{eff}(x,Q^2)$ using several current parametrizations of  
the gluon distributions in the nucleon \cite{MRS,CTEQ,GRV} are presented 
in Fig. 12
for $Q^2$=10 and 40 GeV$^2$ as functions of $x$ (dotted lines) together with 
the experimental ZEUS data point \cite{ZEUS}. One can see that, typically, they 
reduce the $Q^2$ dependence by one power of $Q$ for $Q^2 \sim 10$ GeV$^2$.

A {\it preasymptotic\/} effect  may 
come from higher order corrections to the $1/Q^6$ factor, see, for instance,
the discussion in Sect. IV. The corresponding term, $T(Q^2)$ of
Eq. (\ref{tfactor}) and depicted in Fig. 5, yields an effective power of 
$p_{eff}(Q^2)={\partial \ln T(Q^2) \over \partial\ln Q}$ to the
$Q^2$ dependence of the cross section, where $p_{eff}(10 {\rm GeV}^2) 
\approx 1.5$ and $p_{eff}(40 {\rm GeV}^2) \approx 0.8$.  
Thus, combined, these two effects result in 
\begin{equation}
n_{eff}(x,Q^2)= 6-\left(m_{eff}(x,Q^2)+p_{eff}(Q^2)\right) \ ,
\label{neff2} 
\end{equation}
as shown also in Fig. 12.
One can see from the figure that, in the ZEUS range,
the $n_{eff}$ we find is consistent with the experimental number of ZEUS. 
Note that if we include $Q^2$ rescaling, which was discussed in 
Sect. IV.C, 
the $n_{eff}$'s we find increase by about 0.3 for the kinematics
of the HERA experiment.

\section{Restoration of flavor symmetry }

Longitudinal vector meson production is dominated by small interquark
distances in the vector meson wave function. Therefore the
factorization/decoupling theorem
can be used to calculate the cross section for
hard diffractive processes in QCD without any model assumptions.  For 
$M_V^2 \ll Q^2$, all dependence on the quark masses and thus on flavor
is contained in the light-cone wave function of the vector meson
and not in the scattering amplitude. This prediction is non-trivial,
since experimentally the coherent photoproduction of mesons containing
strange or charm quarks is strongly suppressed as compared to the
SU(4) prediction for the ratio of the production cross sections for
various vector mesons, which is
\begin{equation}
\rho^o : \omega : \phi : J/\Psi =9 : 1 : 2 : 8 \ .
\label{eq21}
\end{equation}
Experimentally, this suppression factor is $\approx 4$ for $\phi$-mesons
and $\approx 25$ for the $J/\Psi$. Thus, QCD predicts a dramatic increase
of the $\phi:\rho^o $ and $J/\Psi:\rho^o$ ratios at large $Q^2$ \cite{Halina}.
We expect that for the $\phi:\rho$-ratio, the pQCD limit would be reached 
rather early, while the restoration of $SU(4)$ in $J/\Psi$-meson production 
would require extremly high $Q^2$. This will be clarified further in the next 
section.
Besides, QCD predicts a slow increase of the relative yield of heavy flavor
production with decreasing $x$ since the virtuality which enters in the
gluon distribution is larger for heavy flavor production.

Moreover, an additional enhancement of heavy flavor production
is expected since, for 
heavy quarkonium states, the probability for a $q$ and $\bar q$ to be  
close together is larger.  In fact, Eq.~(\ref{eq5a}), which was derived from
QCD, predicts for the production ratio of mesons $V_1$ and $V_2$, at
large $Q^2$, that

\begin{equation}
 \left. {\sigma (\gamma ^*_L + T \rightarrow V_1 +T) \over
\sigma (\gamma ^*_L + T \rightarrow V_2 +T)} \right|_{~Q^2
\gg M_{V_1}^2,\,M_{V_2}^2} =~
{M_{V_1}~\Gamma_{V_1 \rightarrow e^+e^-}~\eta_{V_1}^2(Q^2)\over
 M_{V_2}~\Gamma_{V_2 \rightarrow e^+e^-}~\eta_{V_2}^2(Q^2)} \ .
\label{eq22}
\end{equation}
Values of $\eta_V$ for mesons built of light quarks seem to be close to
the asymptotic value of $\eta_V=3$ already for moderate $Q^2 \sim 10$ GeV$^2$.
This reflects the observation of a QCD sum rule analysis \cite{CZ} as well as
lattice QCD \cite{Negele} and the instanton 
liquid model \cite{Shuryak} that the nonperturbative interaction in
the vector channel is weak. 

Based on the measured values of $\Gamma_{V \rightarrow e^+e^-}$ and
estimates of $\eta_{V}$ \cite{CZ} for $\rho$ and $\phi$ and the
charmonium model for $J/\Psi$, we observe that
Eq.~(\ref{eq5a}) predicts a significant enhancement of 
$\phi$ and $J/\Psi$ production as compared to the SU(4) prediction of Eq. 
(\ref{eq21}):
\begin{equation} 
\rho^o : \omega : \phi
: J/\Psi =9 : (1*0.8) : (2*1.0) : (8*1.9) \ ,
\label{eq23}
\end{equation}
At very large 
$Q^2\gg M_V^2$, the $q\bar q$ wave functions of all mesons
should converge to a universal asymptotic wave
function with $\eta_V=3$ \cite{CZ}. In this limit, further enhancement of 
heavy resonance production is expected,
\begin{equation}
\rho^o : \omega : \phi : J/\Psi =9 : (1*0.8) : (2*1.2) : (8*3.5) \ .
\label{eq24}
\end{equation}
It is important to investigate these ratios separately for the
production of longitudinally polarized vector mesons, where hard
physics dominates, and for transversely polarized vector mesons, where
the interplay between soft and hard physics is most important.

\section{Production of excited vector meson states}

Equation~(\ref{eq5a}) is applicable also to the production of
excited vector meson states if their masses $M_{V'}$ satisfy the condition
that $M_{V'}^2\ll Q^2 $.  In this limit, it predicts comparable
production of excited and ground states.  There are no estimates of
$\eta_{V'}$; so, following the above discussion, we assume that the
$q \bar q$ wave functions of a  
$\rho'$, $\omega'$ and $\phi'$ are close to the asymptotic value,
and, as a rough estimate, we will assume that $\eta_{V'}=\eta_V$. Using
the information on the mesons' decay widths from the Review of Particle
Properties~\cite{RPP} and results of an analysis by Clegg and Donnachie 
\cite{CD} of the properties of excited light vector mesons, we find that:
\begin{eqnarray}
\rho(1450)  :\rho^o &\approx& 0.45-0.95     \ , \nonumber \\
\omega(1420):\omega &\approx& 0.46          \ , \nonumber \\
\rho(1700)  :\rho^o &\approx& 0.22 \pm 0.05 \ , \nonumber \\
\omega(1600):\omega &\approx& 0.48          \ , \nonumber \\
\phi(1680)  :   \phi&\approx& 0.85          \ , \nonumber \\
\Psi' :     J/\Psi  &\approx& 0.5           \ .
\label{eq25}
\end{eqnarray}
In view of substantial uncertainties in the experimental widths of
most of the excited states as well as in the values
of $\eta_{V'}$, these numbers can be considered as good to about a
factor of two (except for the $\Psi'$ where $\Gamma_V $ is well
known). 
For example, if we would just sum over the partial widths of the 
$\omega(1600)$, as listed in the Review of Particle Properties \cite{RPP},
we would get $\omega(1600): \omega \ge 0.93$. Note, also, that in quark 
models one would expect $\rho(1450):\rho \approx 
\omega(1420):\omega$ and $\rho(1700):\rho \approx 
\omega(1600):\omega$, which is not satisfied in the cases of the $\rho(1700)$
and $\omega(1600)$.

Note that the essential transverse momenta in the $q \bar q$ wave function 
of the excited states are likely to be significantly larger than in the 
ground state.  This is dictated by the spectrality condition in the 
bound state equation for the minimal Fock $q\bar q$ component
of the light-cone wave function of the vector meson, which ensures that 
\begin{equation}
{m^2 +k^2_t \over z (1-z)} \ge M_V^2 \ .
\end{equation}
The estimate which follows from this equation,
$\sqrt{\left<k_t^2\right>} \approx M_{V'}/2$,
shows that the transverse momenta characteristic for the minimal Fock
component of the wave function should increase with the mass of the excited 
state.  Hence, we made a model estimate for the $Q^2$ dependence of the ratio 
$\rho(1700)\!:\!\rho$ taking model I type wave functions 
for the $k_t$-dependence ($\phi(k_t) \propto {1 \over (k_t^2 +\lambda^2)^2}$
as introduced in Sect. III) with $\sqrt{\left<k_t^2(\rho)\right>}=450$ MeV/c, 
and $\sqrt{\left<k_t^2(\rho)\right> / \left<k_t^2(\rho')\right>}=
M_{\rho} / M_{\rho'}$. The corresponding results for the $Q^2$ dependence
of the ratio $\rho(1700)\!:\!\rho$ are displayed in Fig. 13 as a solid line. 
We also present in this figure the ratios for $\Psi'\!:\!J/\Psi$ and 
$J/\Psi\!:\!\rho$ production, normalized to their asymptotic values at 
$Q^2 \rightarrow \infty$ given in Eqs. (\ref{eq24}) and (\ref{eq25}).


\begin{figure}[htp]
\figa{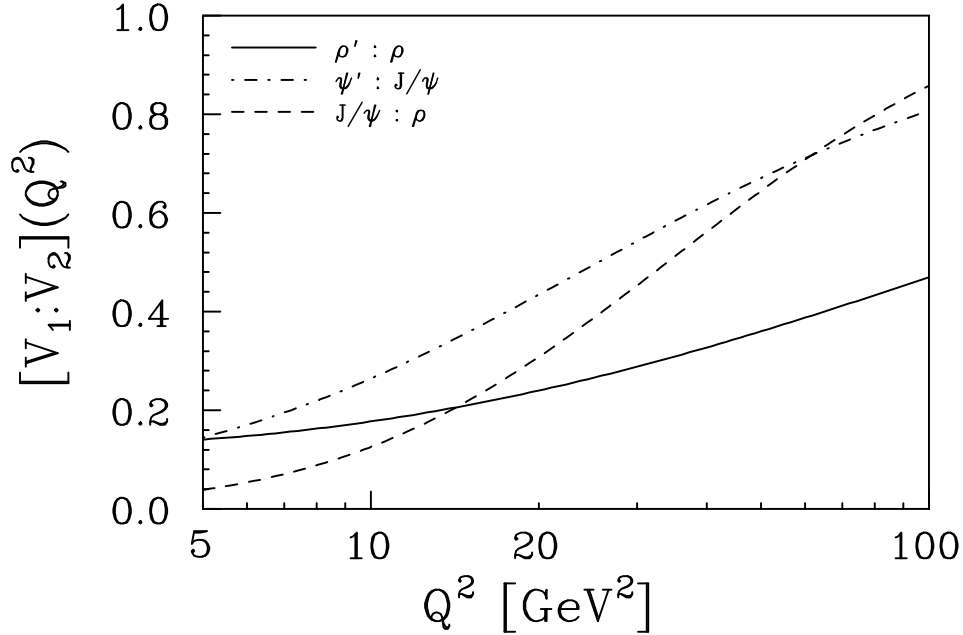}
\caption{Relative yields in the diffractive electroproduction of various vector 
mesons normalized to their asymptotic values at $Q^2 \rightarrow \infty$
given in Eqs. (\protect\ref{eq24}) and (\protect\ref{eq25}).}
\end{figure}

Larger average momenta in the excited mesons' wave functions imply that the
important virtualities in the gluon density essential for a
given $Q^2$ are larger for the production of excited states.
This should manifest 
itself also in an increase of the ratios $V'\!:\!V$ with decreasing $x$,
since the parton distributions depend on $\left<k_t^2(V)\right>$,
and they increase faster at larger $\left<k_t^2(V)\right>$.

For those excited states, which in the nonrelativistic quark model are just 
radial excitations of the corresponding ground state (presumably the 
$\rho(1450)$ and the $\omega(1420)$), and in the $Q^2$ region where one can 
apply perturbative QCD, our predictions for the increase of the $V':V$ ratios 
are qualitatively similar to whose of Nemchik et al. \cite{nemch}, who predict 
an increase of $\rho':\rho$ to 1 at large $Q^2$. At the same time,
in the calculations of Ref. \cite{nemch},
the effects of the quark Fermi motion in the diffractively produced vector 
mesons, which constitute the dominant contribution to the production ratios 
at larger $Q^2$, were not accounted for.
As a result, the saturation of the $V':V$ ratio occurs in the model of, e.g., 
Ref. \cite{nemch} already at $Q^2 \sim 10$ GeV$^2$.

Another important difference stemming from the use of nonrelativistic 
oscillator wave functions, which do not take into account the $Q^2$ evolution 
of the minimal Fock space wave functions (which is a strict consequence of 
QCD \cite{BL}), is the qualitative difference between the behavior of 
radial and orbital (e.g. $D$-state) excitations. The production of the
latter is strongly suppressed at large $Q^2$ in the approximation considered 
in Refs. \cite{kopel2,nemch,benha2}, while we find essentially the same rate 
for both classes of excited states.  So without mentioning this explictly, the 
authors of Refs. \cite{kopel2,nemch,benha2} considered $\rho'$ and $\phi'$
corresponding to radial excitations only.\footnote{We are indepted to 
B.Z. Kopeliovich for clarifying what $\rho'$-state they considered.}  
According 
to the current interpretation of the data, the $\rho(1450)$ and $\omega(1420)$ 
are radial excitations, while the $\rho(1700)$ and $\omega(1600)$ are $D$-wave 
states.  Note that for the cases where the uncertainties in the 
analysis of Clegg and Donnachie \cite{CD} for $\Gamma_{V \rightarrow e^+e^-}$ 
are smaller - $\omega(1420)$ and $\omega(1600)$ - we find that the expected 
yields at large $Q^2$ as given by Eq. (\ref{eq25}) are comparable.

To summarize, a substantial production of excited resonance states is
expected at large $Q^2$ at HERA. A measurement of these reactions may
help to better understand the dynamics of the diffractive production
of vector mesons
as well as the light-cone minimal Fock state wave functions of these
excited states.  It may provide the first {\it three dimensional\/} images of
these states, and it would also allow to look for the second missing
excited $\phi$ state, which is likely to have a mass of about 1900 MeV
to follow the pattern of the $\rho$, $\omega$ and $J/\Psi$ families.

The relative yield of excited states produced by virtual photons is
expected to be higher than that for real photons since the Vector Dominance
Model (VDM) \cite{bauer} together with Eq.~(\ref{eq5a}) leads to
\begin{eqnarray}
\left. {\sigma (\gamma  + N \rightarrow V +N) \over
\sigma (\gamma   + N \rightarrow V' +N)}
 {\sigma (\gamma ^*_L + N \rightarrow V' +N) \over
\sigma (\gamma ^*_L + N \rightarrow V  +N)}
\right|_{~Q^2 \gg M_{V'}^2,M_{V}^2}
=&&\nonumber \\
={M_{V'}^2 \over M_V ^2}~{\eta_{V'}^2(Q^2)\over \eta_{V}^2(Q^2)}&&
\ge {M_{V'}^2 \over M_V ^2}  \ .
\label{eq26}
\end{eqnarray}
In the last step, we used the empirical observation that for the effective
cross sections of $V'N$ and $VN$ interactions, which enter in the VDM
model,\footnote{Note that these effective cross sections have no
  direct relation to the genuine interaction cross sections.  For
  example, based on geometrical scaling, one expects the interaction
  cross section to increase with the size of the projectile
  approximately as $R=\sigma_{tot} (\Psi'N)/\sigma_{tot}(J/\Psi N)
  \sim R^2_{\Psi'}/R^2_{J/\Psi} \sim 4$.  However, if one applies 
  the VDM to the extraction of cross sections from
  photoproduction of $J/\Psi$ and $\Psi'$, one finds $R \sim 0.7 \sim
  M^2_{J/\Psi}/M^2_{\Psi'}$.  This trend seems to reflect effects of
  color screening in the production of heavy quarkonium
  states~\cite{FS85,FS88}.  Note, also, that photoproduction data do not
  resolve the $\rho(1430)$ and $\rho(1700)$.  In the case of $\rho'$
  photoproduction off nuclei, similar nuclear absorption effects are
  observed for the production of the $\rho$ and $\rho'$, indicating
  $\sigma(\rho'N) \approx \sigma(\rho N)$.  At the same time,
  application of the VDM to the process $\gamma~p \rightarrow \rho' p$ 
  leads to $\sigma(\rho'N) \approx (0.37 \pm 0.07)\,\sigma(\rho N)$, where
  we used the analysis of Ref. \cite{CD} of the $V \rightarrow e^+e^-$
  widths and the data of Chapin et al. \cite{chapin} for diffractive 
  $\rho'$ production.
  The observed pattern indicates that the production of
  $\rho'$ is dominated by average quark-gluon configurations (large
  absorption cross section), while the probability of these
  transitions is suppressed since the transition $\gamma^*\rightarrow
  V$ emphasizes the role of small configurations.
  Another possible interpretation (which maybe complementary to the one above)
  is a large contribution of nondiagonal transitions $\rho' \rightarrow \rho$ 
  in the scattering off nuclei. CT arguments \cite{FS91} (p.170) suggest that 
  $V\rightarrow V'$ transitions could indeed be very appreciable.}
${\sigma_{tot}(V'N) \over \sigma_{tot}(VN)} \leq 1$, and that $\eta_V$ and 
$\eta_{V'}$ are close to their asymptotic values for light mesons while 
for heavy quark systems the values of $\eta_V'$ should be 
closer to the static limit of $\eta_V=2$.

Equation (\ref{eq5a}) is applicable also to vector meson
production in weak processes.  Consider, for example, the diffractive
production of a $D^{*\pm}_s=(c\bar s)$ meson in $ W^{\pm}N$ scattering. To
calculate this cross section, it is sufficient to substitute in
Eq.~(\ref{eq5a}) the electromagnetic coupling constant with
$g \cos\theta_C$, where $\theta_C$ is the Cabibbo angle. 

It is often discussed that the production of hadrons in $e^+e^-$annihilation 
at intermediate masses ($M_h \le 2 \div 2.5$ GeV) can be described as the 
production 
and decay of several vector meson resonances. Within this approximation, 
we can generalize Eq. (\ref{eq5a}) to the case of diffractive
electroproduction  of hadron  states in the continuum. To deduce the relevant 
formulae,
we assume that for excited vector meson resonances $\eta_V$ is close to its 
asymptoic value, $\eta_V $ = 3. This assumption really follows from
the application of the
dispersion sum rule approach. In the limits 
$M^2_h \le 5 {\rm GeV}^2 \ll Q^2$, $x \ll 1$ and $ Q^2 \rightarrow \infty$, 
we obtain:
\begin{equation}
\left. {d\sigma^L_{\gamma^*N\rightarrow hN}\over dtdM^2}\right|_{t=0} =
{9 T^2(Q^2)M_h^2 \sigma(e^{+}e^- \rightarrow  h) \alpha_s^2(Q)\left|
\left(1 + i{\pi\over2}{d \over d\ln x}\right) xG_T(x,Q)\right|^2 \over 4 
N^2_c\alpha_{EM}Q^6} \ .
\label{mpord}
\end{equation}

In is worth emphasizing that in the case of diffraction to high enough masses,
where the intermediate $\gamma^*$ state can be approximated by a free $q 
\bar q$ pair, the contribution of large transverse distances is not 
suppressed and the cross section
gets both soft and hard contributions. Consequently, in the intermediate 
$M_h^2$ range, where resonances are still present but the continuum is already
essential, one should expect a faster increase with energy (at fixed $Q^2$)
of the resonance contribution.

To summarize, the investigation of exclusive diffractive processes
appears to be the most effective method to measure the minimal Fock state
$q\bar q$ component of the wave functions of vector mesons as well as the
light-cone wave functions of any small mass hadron system having
angular momentum one.  It seems that the cross sections of these processes
are rather sensitive to the three dimensional distribution of color within 
the wave functions of vector mesons. This observation could be very helpful 
in expanding methods of lattice QCD into the domain of high energy processes.

\section{The kinematical boundary for the region of applicability of the
         leading logarithm approximations to QCD}

        In the above, we have explained that the increase of the cross section 
with energy (${1 \over x}$) at fixed $Q^2$ originates from the energy 
dependence of the $q \bar q T$ cross section, see Eq. (\ref{eq7}) and Fig. 3.
But, in the region dominated by hard physics,
the value of this cross section is restricted by the unitarity 
of the $S$-matrix,  
and even more 
stringent 
restrictions follow 
from the condition that the leading twist term 
should be significantly larger than the next to leading twist term.
In detail, let us consider the scattering of a small object,
a $q\bar q$ pair, from a large object, a nucleon. If only hard physics
were relevant for the increase of the parton distributions at small $x$,
the radius of the $q \bar q$-nucleon interaction should 
be practically independent on energy, because, in this case, Gribov diffusion 
in the impact parameter space would be determined by the scale characteristic 
for hard processes.  We shall explore this observation to establish
the region of applicability of the leading
$\ln{1\over x}$ and $\ln{Q^2\over\Lambda_{QCD}^2}$ approximations.

        Thus, the geometrical restriction is that,
in the hard regime,
the inelastic 
cross section cannot exceed the geometrical size of the target nucleon. 
An equivalent, but practically more convenient, formulation of
this important property of hard processes is that the slope of the differential 
cross section should not increase with energy. It is just this property of hard 
processes which gives the possibility to deduce an unitarity limit for their
cross sections and/or on the region dominated by hard physics.

        We will start our considerations with the derivation of the boundary
for the region where the decomposition over powers of ${1\over Q^2}$ can be
a sensible approximation.  A formal method to deduce this restriction
is to consider scattering of heavy quarkonium $Q \bar Q$ states from a hadron 
target, and to vary the mass of the heavy quark.  Since the result of such a 
derivation is evident, in this paper we use an equivalent but
shorter way to deduce this inequality \cite{Halina}: We establish the limit by
applying the optical theorem to calculate the elastic cross section for  
scattering of a small-transverse-size $q\bar q$ pair
off a nucleon: 
\begin{equation}
{d\sigma_{el}\over dt}(q \bar qN) ={\sigma_{tot}^2\over 16\pi }\,e^{Bt}\,
(1 +\beta^2) \ ,
\label{61d}
\end{equation}
where 
\begin{equation}
\beta= {  Re A_{q \bar qN}/Im A_{q \bar qN}} \simeq {\pi \over 2}
{d \ln(xG_N(x,Q^2)) \over d\ln x} \ .
\end{equation}
For simplicity, we parametrize the dependence of the differential cross 
section on the momentum transfer $t$, at small $t$ (which dominate the 
cross section), in the form of an exponential. 
As the observed $t$ dependence of hard 
diffractive processes is close to that given by the square of the nucleon's
electromagnetic form factor, i.e., $\sim \exp({r_N^2\over 3} t)$, 
to visualize the deduced limit, it is convenient to express the
experimental value of the slope of the $t$ dependence of the cross section
through the average quadratic radius of a nucleon: $B \approx 4 \div 5
{\rm~GeV}^{-2} \approx {r_N^2 \over 4}$ where $r_N=0.8$ fm.
It follows from QCD (cf. discussion in Sect. III.A) that this value should
be universal for all hard processes.  Existing data on hard diffractive 
production of vector mesons are in line with this QCD prediction, 
and we can use it to calculate the elastic cross section:
\begin{equation}
\sigma_{el} ={\sigma_{tot}^2\over 16\pi B } (1 + \beta^2)
\approx {\sigma_{tot}^2 \over 4 \pi r_N^2} (1 + \beta^2) \ .
\label{61ee}
\end{equation}
In the derivation of the boundary for the region where the leading
$\ln{1\over x}$ and $\ln{Q^2\over\Lambda_{QCD}^2}$ approximations are
applicable, one should actually add to Eq. (\ref{61ee}) 
the contribution from proton dissociation, i.e., one should multiply 
$\sigma_{el}$ by $(1+\gamma)$ where $\gamma \sim 10 \div 30$\%.

The necessary condition for the applicability of the decomposition of the deep 
inelastic cross section over powers of $Q^2$ is that the leading twist
contribution should exceed nonleading twist terms. In our case, this 
implies that $\sigma_{inel} \gg \sigma_{el}(1+\gamma)$, where
$\gamma$ accounts for the dissociation of the proton.  So, the boundary  
for the applicability of the decomposition over powers of $1/Q^2$ is 
${\sigma_{el}(1+\gamma)\over \sigma_{inel}}=y \ll 1$.
Using Eq. (\ref{61ee}) and above discussed value of $B$, we obtain:
\begin{equation}
\sigma_{inel} \ll {\pi {r_N}^2 \over  (1 + \beta^2)}
                  {4y\over (1 + \gamma)(1+y)^2} \ .
\label{61e}
\end{equation}

Thus the region of applicability of the QCD evolution equation is: 
\begin{equation}
\sigma_{inel}(q \bar q N)~=~{{\pi}^2 \over 3} b^2 \,
\alpha_s\!\left({9.2\over b^2}\right) 
xG_p
\!\left(x,{9.2\over b^2}\right)~
\ll {\pi {r_N}^2 \over  (1 + \beta^2)}{4y\over (1+\gamma)(1+y)^2}
\label{61f}
\end{equation}
The unitarity of the $S$-matrix, which corresponds to the 
condition $y=1$, was discussed in Refs. 
\cite {nikol4,baron,Halina}.
It can only be translated into an actual kinematical boundary if one 
considers purely hard physics, where $B$ is restricted 
by the nucleon size.  In general, however,
the slope $B$ and its energy dependence
have to be evaluated separately, which has not been done so far.
Such analysis would lead to the Froissart limit, with $B$ and the cross 
section slowly increasing with energy.
Note also that Eq. (\ref{eq7}) has been deduced in the approximation where
only the leading power of $b^2$ has been taken into account.
A more stringent limit follows from the requirement that the
single particle density in the center of rapidity should be positive.
Really, it follows from the application of 
Abramovski$\breve{{\rm i}}$, Gribov and Kancheli (AGK) combinatorics
\cite{AGK} that this multiplicity is proportional to $\sigma_{inel}
-4(\sigma_{el}+\sigma_{diff})$. Thus, in the approximation when only
one hard rescattering is accounted for, we have $y \le 1/4$.

Therefore, the kinematical boundary for the region of applicability
of the leading logarithm approximation is significantly more stringent
than that given by the unitarity condition:
\begin{equation}
\sigma_{inel}(q \bar q N)~=~{{\pi}^2 \over 3} b^2 \,
\alpha_s\!\left({9.2\over b^2}\right)xG_p\!\left(x,{9.2\over b^2}\right)~
\ll {16 \over 25}\,{\pi {r_N}^2 \over  (1 + \beta^2)(1+\gamma)}
\label{61g}
\end{equation}
Eq. (\ref{61g}) 
allows us to estimate down to what $x \equiv x_{lim.}$
the equations for $\sigma_L$ and  $\sigma_{\gamma^*_LN \rightarrow 
\rho N}$ can be applied.  For certainty, we choose $\gamma=0.2$ 
in our evaluation.
Results of a calculation of $x_{lim.}(Q^2)$ based on Eq. (\ref{61g})
for the two most current gluon parametrizations are presented in Fig. 14. 
The boundary is even more stringent for $\rho$-meson 
production since the average $b(Q)$ are larger in this case (see the dashed 
and dotted curves in Fig. 14).  Actually, it is likely that the slow down 
starts earlier, in particular due to dispersion in $b$.
However, a detailed analysis of this effect is beyond the scope 
of this work.


\begin{figure}[htp]
\figa{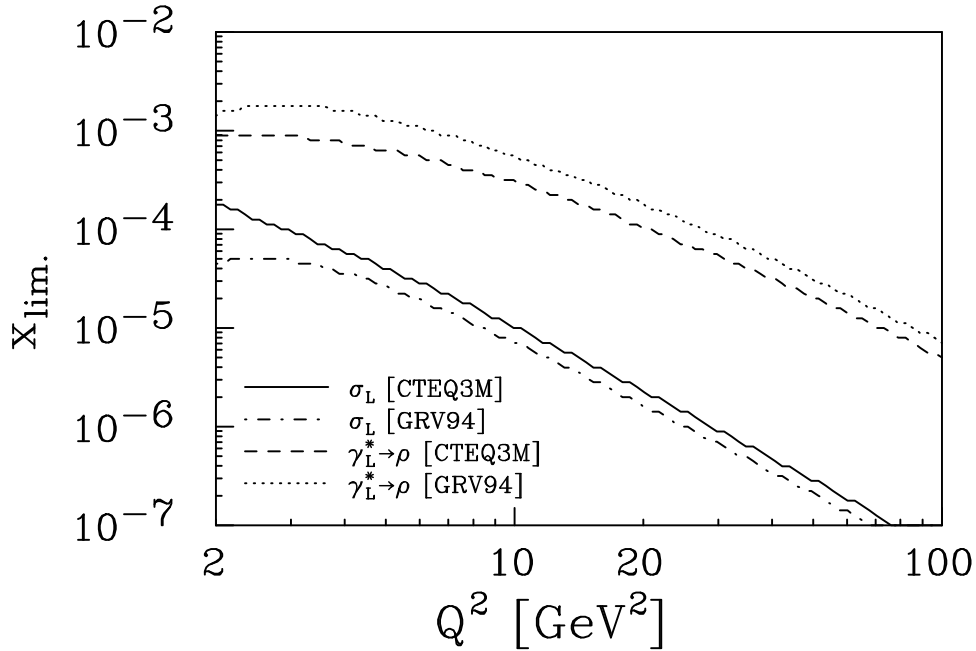}
\caption{The boundary of the region of applicability of pQCD for
a nucleon target, $x_{lim.}(Q)$, as determined 
from Eq. (\protect\ref{61g}) for deep inelastic scattering ($\sigma_L$) as 
well as diffractive electroproduction of $\rho$-mesons 
($\gamma^*_L N \rightarrow \rho N$).}
\end{figure}

The advantage of the approach used in this paper, as compared to previous 
attempts (for a review and references see Refs. \cite{BKCK,Levin}),
is that we actually deduce the QCD
formulae for the cross section for scattering of a $q\bar q$ pair off a
hadron target.  For this cross section, the geometrical limit (including
numerical coefficients) unambiguously follows from the unitarity of the 
$S$-matrix, that is from the geometry of the collision 
(under the assumptions made in  Refs. \cite{BKCK,Levin} on the dominance of 
hard physics near the unitarity limit).  As a result, we obtain a
more stringent inequality which contains no free parameters.  Recently, a 
quantitative estimate
of the saturation limit was obtained~\cite{Collins} by considering the
GLR model~\cite{LR,MQ} of nonlinear effects in the parton evolution
and requiring that the nonlinear term should be smaller than the
linear term.  The constraint obtained in that paper is 
numerically much less restrictive than our result: 
the parameter characterizing the scale of corrections found in Ref. 
\cite{Collins}
differs from ours by a numerical factor $\sim {1 \over 16}$. 
The
qualitative reason for this difference is that the GLR model neglects 
elastic rescatterings of the $q \bar q$ pair,
and it does not take into account the requirement of positiveness
of the single particle density.
The consideration of a specific cut of 
the double scattering diagram, which corresponds to the 
shadowing of the single hadron multiplicity, 
and the requirement that the single multiplicity should be 
positive leads, within this model, to the restriction
that 
$\sigma_{el}+\sigma_{diff}$ 
cannot exceed $1/4\,\sigma_{inel}$. 
The numerical coefficient follows then from the application of 
Abramovski$\breve{{\rm i}}$, Gribov and Kancheli (AGK) combinatorics
\cite{AGK}. 
This observation indicates that, in the GLR model
\cite{LR,MQ}, hard double scattering may decrease
the parton density by not more than $25\%$ 
(and by no more than the observed diffraction fraction of the cross 
section which is $\sim 15\%$ at HERA).  As usual,
if the corrections become large enough, one cannot restrict the consideration 
to one particular contribution. 

We want to point out that the deduced constraint
implies only a restriction on the limiting behavior of the cross 
sections for hard processes, but it gives no clues whether this boundary
can be achieved.
In reality, the increase of the cross sections of hard processes
should slow down already at larger $x$ than those given by the constraint
which we deduced here.
This is because in the derivation of the boundary we ignore, amongst others, 
the for Feynman diagrams characteristic diffusion in the parton ladder to 
large distances, 
effects related to the dispersion of $b$ as well as other hard 
diffractive processes which also fastly increase at $x \rightarrow 0$.
The dynamical mechanism responsible for the slowing down of the increase of 
the parton distributions, so that they satisfy, for instance, Eq. (\ref{61e}),
is subject of discussions.  In particular,
the triple Pomeron mechanism for shadowing suggested in Ref. \cite{LR} does
not lead to large effects at HERA energies, especially if one assumes a
homogeneous transverse density of gluons~\cite{Levin,Martin}.

\section{Limits of applicability of the QCD evolution equations to the 
         scattering off heavy nuclei}

The application of similar consideration to the scattering 
off nuclei leads to a more severe restriction on the
gluon distribution in nuclei:  
\begin{equation}
\sigma_{inel}(q \bar q A) \le  {{\pi} R_A^2 \over (1 + \beta^2)}
                               {4y\over (1 + \gamma_A)(1+y)^2} \ .
\label{61ff}
\end{equation}
Here, $R_A$ is the radius of the nucleus and $\gamma_A$ stems from the
dissociation of the nucleus. 
$\beta=Re A_{q \bar qN}/Im A_{q \bar qN}$ is determined from the $x$ 
dependence of the gluon density
and should be universal within a considered approximation.
For a heavy nucleus, we may safely substitute $y=1$ since the total
and elastic cross sections are then not far from the black body limit,
where $\sigma_{tot}^A \approx 2\pi R_A^2$ and $\sigma_{el}^A \approx
\pi R_A^2$, even for moderate values of $\sigma_{inel}(q \bar qN)$.
In the case of a nuclear target, the limit we consider here is
worth a separate investigation since interesting effects may arise
in the kinematics where the parton densities of a nucleon
are still far from the unitarity limit. A.Mueller and J.Qiu \cite{MQ}
were the first to draw attention to nuclear enhancement of nonlinear 
effects in the QCD evolution equation.

The inequality of Eq. (\ref{61ff}) corresponds to a rather 
stringent restriction on the gluon distribution in heavy nuclei:
\begin{equation}
{1 \over A}
xG_A(x,Q^2) \le {3 r_0^2 \over \pi \alpha_s A^{1/3}}~{Q^2 \over 9.2}~
{1 \over  1 + \beta^2}~{4y\over (1 + \gamma_A)(1+y)^2} \ ,
\end{equation}
where $R_A=r_0 A^{1/3}$ and $r_0 \approx 1.1$ fm. For $Q^2= 10{\rm~GeV^2}$
$\alpha_S=0.3$, 
$\gamma_A=0$, and $y=1$,
this yields
\begin{equation}
{1 \over A} xG_A(x,Q^2=10{\rm~GeV^2}) \le {75 \over A^{1/3}} \ .
\label{AA}
\end{equation}
A more accurate estimate for this constraint could be 
obtained by using the
discussed inequality for each $b$ and then averaging over $b$ but not just
taking the value in the mean point.

In the above, to quantify the kinematical bounds for $\sigma_L$ and vector 
meson production, we neglected  the dispersion 
over $b$ in the corresponding amplitudes.
But, in practice, we need the amplitude in $k_t$-space, i.e., the Fourier
transform of the cross section which we discussed above with weights given 
by the wave functions of the photon or vector meson, respectively. 
For a more accurate estimate, we need to apply the kinematical limit 
for fixed $b$ and then calculate the cross sections based on Eqs. (\ref{brep})
and (\ref{sigmal}) with $\sigma_{q \bar q T}$ satisfying the bound.

Let us restrict our analysis here to the case of heavy nuclei as targets,
where these effects become important at significantly larger $x$ than for
a nucleon target, and where uncertainties in the estimate of the 
region of applicability of the evolution
equation are comparatively small. In fact, we will focus our analysis 
on the region  which
would be accessible for HERA with nuclear beams and for the LHC: 
$x \in 10^{-3} \div 10^{-4}$. In this kinematics, the gluon distribution
in the nucleon is already known reasonably 
well from the recent HERA experiments.

To estimate nonlinear effects, we
allow $\sigma_{q \bar q A}(x,b)$ to reach
the unitarity bound, i.e., we set $y=1$ in Eq. (\ref{61ff}),
and if  $\sigma_{q \bar q A}(x,b) \ge \pi R^2_A$,
we substitute it by $ \pi R^2_A$.  Since in the discussed
kinematics, the unitarity corrections to $\sigma_{q \bar q N}$ are still small,
the ratio of cross sections calculated with and without this unitarity 
correction characterizes 
the effective gluon shadowing for $\sigma_L$ and vector meson
production. In the case of $\rho$-meson production,
we chose the same wave function as in Sect. III
with $\sqrt{\left<k^2_t\right>}
=450$ MeV/c. We also checked that varying $\sqrt{\left<k^2_t\right>}$ in 
the range $300 \div 600$ MeV/c practically does not change our results.
The relevant ratios,  
$R_L(A,x,Q^2)={1 \over A} {\sigma^{\gamma^*_L A} \over  \sigma^{\gamma^*_L N}}$ 
and
$R_V(A,x,Q^2)=\left. {1\over A^2} {d\sigma^{\gamma^*_L +A \rightarrow V +A}/dt 
\over d\sigma^{\gamma^*_L +N \rightarrow V +N}/dt}\right|_{t=0}$ 
are presented in Fig. 15 for $A=40$ and $A=238$ and for $Q^2=$2, 5, 
10, 20 and 40 GeV$^2$.  For the sake of an easier
comparison of the shadowing effects, in the case of $\rho$-meson 
production, we present plots for $\sqrt{R_V(A,x,Q^2)}$.
Clearly this estimate is an upper bound for these ratios since it neglects 
absorbtive effects for the case when the interaction  is not black.


\begin{figure}[htp]
\figc{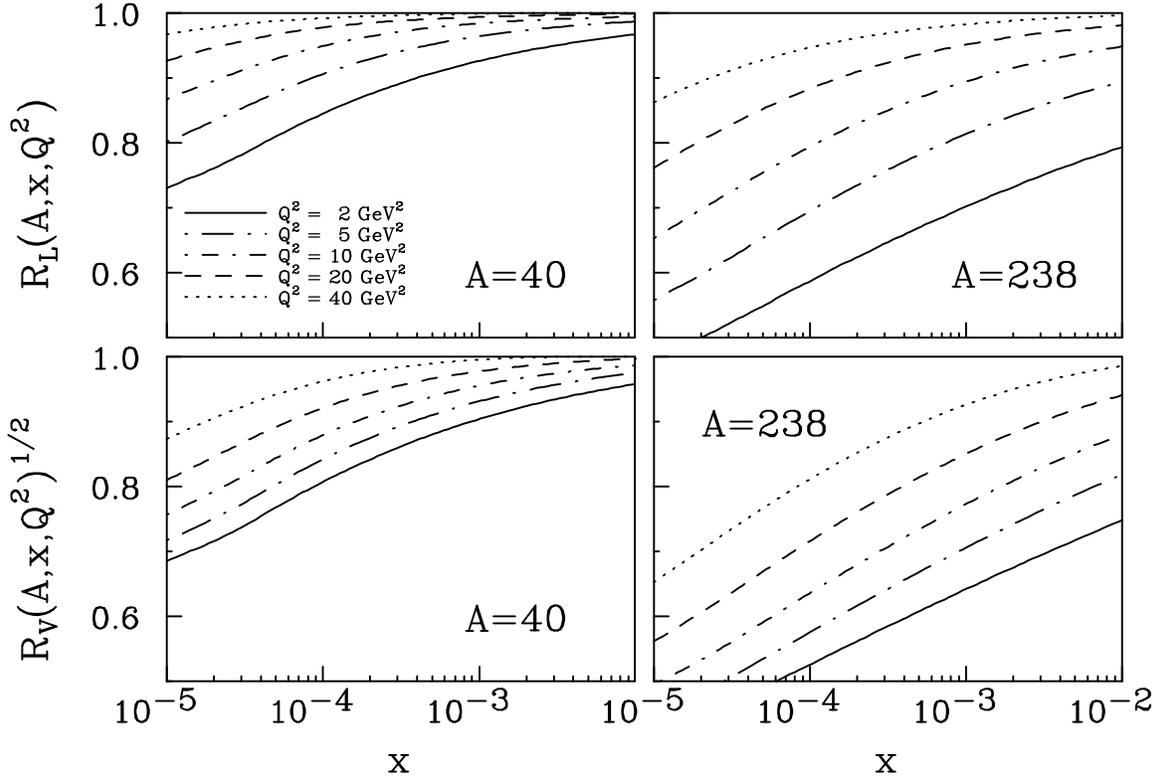}
\caption{Nuclear shadowing in DIS in the gluon channel and in diffractive 
electroproduction of vector mesons is evaluated by employing the nuclear 
unitarity limit.} 
\end{figure}

To estimate nuclear shadowing within a dynamical model, we also consider 
the $b$-space eikonal approximation for the calculation of the same 
quantities. We first calculate the total cross section for the $q \bar q N$ 
interaction by means of the optical theorem,
\begin{equation}
\sigma_{inel} + {\sigma_{tot}^2 \over 16 \pi B}(1 +\beta^2)  
(1+\gamma_A) = \sigma_{tot} \ ,
\end{equation}
and then evaluate the cross section of the $q \bar q A$ interaction
for a given impact parameter within the $b$-space eikonal approximation
using the familiar Glauber-Gribov formula \cite{glauber,gribov}:
\begin{eqnarray}
R_L(A,x,Q^2) &=& 
{1 \over A}
{Re\left[
\int\!d^2b~dz~|\psi_{\gamma^*_L}(b,z)|^2~
\int\!d^2B\left(2-2e^{-(1-i\beta){\sigma_{tot}(b) T(B) \over 2}}\right)
\right]
\over
\int\!d^2b~dz~|\psi_{\gamma^*_L}(b,z)|^2~\sigma_{tot}(b)} \ ,
\label{rhoa}
\\
R_V(A,x,Q^2) &=&
{1 \over A^2}
\left|{
\int\!d^2b~dz~\psi_{\gamma^*_L}(b,z)~\psi_{\rho}(b,z)~
\int\!d^2B\left(2-2e^{-(1-i\beta){\sigma_{tot}(b) T(B) \over 2}}\right)
\over
\int\!d^2b~dz~\psi_{\gamma^*_L}(b,z)~\psi_{\rho}(b,z)~\sigma_{tot}(b)
(1-i\beta)}\right|^2 \ ,
\label{rhob}
\end{eqnarray}
Here, $T(B)= \int_{-\infty}^{+\infty} dz \rho(z,B)$ is the optical thickness
of the nucleus and $\rho(z,B)$ is the nuclear density normalized
so that $\int_{-\infty}^{+\infty} dz \, d^2\!B \, \rho(z,\vec B)=A$.
We use Eqs. (\ref{rhoa}) and (\ref{rhob}) to estimate the nuclear
shadowing for $\sigma_L$ and for $\rho$-meson production employing a 
realistic nuclear density distribution, $\rho(r)\sim 1/(1+\exp[(r-R_A)/a])$,
where $R_A=1.1\,A^{1/3}$ fm and $a=0.56$ fm.  Corresponding results are 
presented in Fig. 16.  


\begin{figure}[htp]
\figc{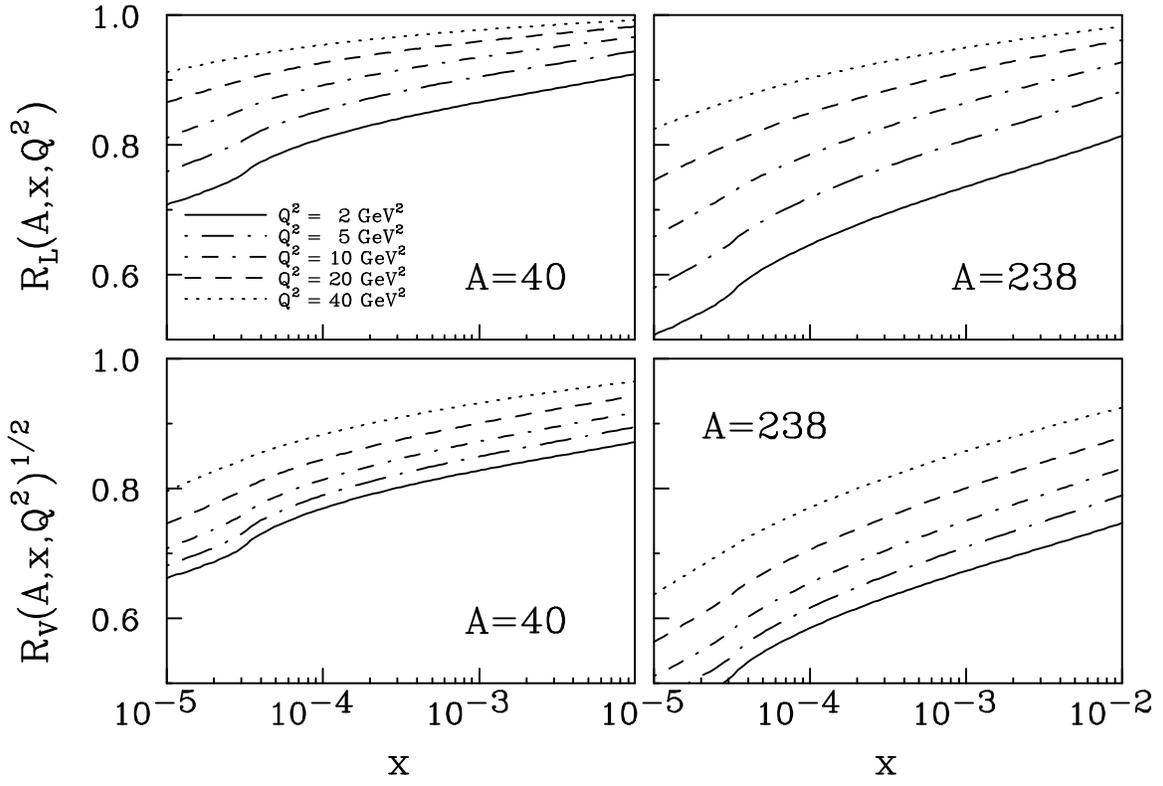}
\caption{Nuclear shadowing in DIS in the gluon channel and in diffractive 
electroproduction of vector mesons is evaluated in $b$-space eikonal
approximation.}
\end{figure}

One can see from Figs. 15 and 16 that significant gluon
shadowing is expected at small $x$, and that shadowing is larger in the 
$\rho$-meson channel, i.e., $\sqrt{R_V(A,x,Q^2)} < R_L(A,x,Q^2)$. This is again
due to the larger $b$ for $\rho$-meson production --
see the discussion in Sect. III.  Also, in line with expectations of
the color transparency logic,
at fixed large $Q^2$, shadowing increases 
with decreasing $x$ -- i.e., $R_L, R_V \ll 1$ for $x \rightarrow 0$ and color
transparency {\it disappears\/} for small enough $x$ -- while, for fixed small
$x$, shadowing decreases with increasing $Q^2$ -- i.e., $R_L, R_V \rightarrow 1$
at large $Q^2$ and color transparency {\it appears\/} for large enough $Q^2$.

Furthermore, from Figs. 15 and 16, we observe that
the nuclear shadowing which we deduce from the unitarity limit and from the
eikonal model is significantly smaller than the black disk limit.
This is partly due to the diffuse edge of the nucleus, and
a practical conclusion from this observation is that nuclear shadowing of the
parton distributions should be considerably larger for central $AA$ collisions 
where the effects of the nuclear surface should be insignificant.

Note, however, that even those estimates are still likely to
underestimate the relevant shadowing since they do
not explicitly include gluon shadowing in the leading twist, i.e. 
logarithmically decreasing at large $Q^2$. 
Still, our estimate corresponds to significantly larger gluon
shadowing than that obtained in a number of previous papers, 
where it was assumed that for $Q^2 \approx$  few GeV$^2$  
the leading twist 
shadowing in the gluon channel and 
for $F_{2A}$ are approximately the same \cite{Q,FS88,FS,Es,EQW}. Note that 
qualitative arguments in favor of larger shadowing in the gluon channel 
were presented in Ref. \cite{FLS93}. The quantities $R_L$ and $R_V$ of
Eqs. (\ref{rhoa}) and (\ref{rhob}) were also evaluated in Refs. 
\cite{kopel2,nikol3} using the two-gluon-exchange model, in which the 
elementary cross section of the interactions with the constituent quarks
does not depend on energy. As a result, in this model, attenuation does not 
depend on $x$ for small $x$.  

In principle, shadowing effects in the kinematics beyond the region of 
applicability of the evolution equation may change the dependence of the
differential cross section on $t$, the momentum transferred to the nucleus
\cite{progress2}. In that realm,
the eikonal approximation leads to the famous pattern with minima and maxima.  
Another well known feature of the eikonal approximation, which is specific for 
potentials fastly increasing with energy, is the shrinking of the diffractive
peak with energy.  Since, within this
model, both soft and hard physics are important, the radius of a nucleon may 
increase with energy, and, for this regime, the unitarity restriction for 
processes dominated by hard physics, which we discussed above,
could be exceeded and the cross section may increase with $x$, but more 
slowly as a power of $\ln x$. For a nuclear target, such effects are expected 
to be insignificant, although for a nucleon target they may be large.

To summarize, the analysis presented in this Section demonstrates that large 
shadowing effects should be present at small $x$ in the gluon channel.
This has a number of implications for high-energy nucleus-nucleus
collisions, where the predictions for hadron production strongly 
depend on gluon shadowing, as was pointed out, for example, in 
Ref. \cite{GW}.

\acknowledgments{We would like to thank H. Abramowicz, J. Bartels, J. Bjorken, 
S. Brodsky, A. Caldwell, J. Collins, V. Gribov, and A. Mueller for fruitful 
discussions on the interplay of soft and hard physics and of the methods of 
their investigation.  
This work was supported in part by the
Israel-USA Binational Science Foundation Grant No.~9200126, by the
MINERVA Foundation of the Federal Republic of Germany, and by the
U.S.~Department of Energy under Contract No.~DE-FG02-93ER40771.
Two  of us (W.K. \& M.S.)  thank the DOE's Institute for Nuclear Theory at
the University of Washington for its hospitality and support during 
the workshop ``Quark and Gluon Structure of Nucleons and Nuclei''.
}

\end{document}